\documentclass[aps,nofootinbib,prr,reprint,superscriptaddress,longbibliography]{revtex4-1}
\usepackage{amsmath}
\usepackage{graphicx}
\usepackage{amsfonts}
\usepackage{color}
\usepackage{upgreek}
\usepackage{bm}
\usepackage{lipsum}
\usepackage{comment}
\usepackage{enumerate}
\usepackage{dsfont}

\usepackage[table]{xcolor}
\usepackage{textcomp}
\usepackage{amsmath}
\usepackage{amssymb}
\usepackage{graphicx} 
\usepackage{pgf}
\usepackage{epstopdf}
\usepackage{braket}
\usepackage{mathtools}
\usepackage{times}

\usepackage[retainorgcmds]{IEEEtrantools}
\newcommand{\bea}{\begin{eqnarray}}
\newcommand{\eea}{\end{eqnarray}}

\newcommand{\Tr}{\mathrm{Tr}}

\usepackage[colorlinks,linkcolor=blue,urlcolor=blue,citecolor=blue]{hyperref}

\begin{document}

\title{Mixed-state phase transitions in spin-Holstein models}

\author{Brett Min$^*$}
\email{brett.min@mail.utoronto.ca}
\affiliation{Department of Physics and Centre for Quantum Information and Quantum Control, University of Toronto, 60 Saint George St., Toronto, Ontario, M5S 1A7, Canada}
\affiliation{Department of Physics, The University of Tokyo, 7-3-1 Hongo, Bunkyo-ku,
Tokyo 113-0033, Japan}

\author{Yuxuan Zhang$^*$}
\email{quantum.zhang@utoronto.ca}
\affiliation{Department of Physics and Centre for Quantum Information and Quantum Control, University of Toronto, 60 Saint George St., Toronto, Ontario, M5S 1A7, Canada}
\affiliation{Vector Institute, W1140-108 College Street, Schwartz Reisman Innovation Campus Toronto,
Ontario M5G 0C6, Canada}

\author{Yuxuan Guo$^*$}
\email{yuxguo2024@g.ecc.u-tokyo.ac.jp}
\affiliation{Department of Physics, The University of Tokyo, 7-3-1 Hongo, Bunkyo-ku,
Tokyo 113-0033, Japan}

\author{Dvira Segal}
\email{dvira.segal@utoronto.ca}
\affiliation{Department of Chemistry
University of Toronto, 80 Saint George St., Toronto, Ontario, M5S 3H6, Canada}
\affiliation{Department of Physics and Centre for Quantum Information and Quantum Control, University of Toronto, 60 Saint George St., Toronto, Ontario, M5S 1A7, Canada}

\author{Yuto Ashida}
\email{ashida@phys.s.u-tokyo.ac.jp}
\affiliation{Department of Physics, The University of Tokyo, 7-3-1 Hongo, Bunkyo-ku,
Tokyo 113-0033, Japan}
\affiliation{Institute for Physics of Intelligence, The University of Tokyo, 7-3-1 Hongo,
Tokyo 113-0033, Japan}

\def\thefootnote{$*$}\footnotetext{These authors contributed equally to this work. }

\begin{abstract}
Understanding coupled electron-phonon systems is one of the fundamental issues in strongly correlated systems. 
In this work, we aim to extend the notion of mixed-state phases to the realm of coupled electron/spin-phonon systems. Specifically, we consider a two-dimensional cluster Hamiltonian locally coupled to a set of single bosonic modes with arbitrary coupling strength. First, we adopt a pure-state framework and examine whether a ground state phase transition out of the symmetry-protected topological phase can be captured using the standard polaron unitary transformation. This approach involves restricting the analysis to the low-energy manifold of the phonon degrees of freedom. We find that the pure-state approach fails to detect the anticipated transition to a topologically trivial phase at strong spin-phonon coupling. Next, we turn to a mixed-state picture. Here,
we analyze mixed states of the model obtained by tracing out the phonons degrees of freedom. We employ two distinct diagnostics for mixed-state phase transitions: (i) the von Neumann conditional mutual information (CMI) and (ii) the Rényi-2 CMI. 
We argue that both measures detect signatures of mixed-state phase transitions, albeit at different critical spin-phonon coupling strengths, corresponding to subtly distinct notions of the mixed-state phases.
\end{abstract}
\maketitle

\date{\today}

\section{Introduction}
Electron/spin-phonon coupling is one of the most fundamental interactions in quantum matter, influencing the phases of many-body systems. Allowing strongly correlated electron systems to interact with phonon modes at arbitrary coupling strengths gives rise to rich phase diagrams~\cite{Shneyder_2020,Xing_2023}. The standard approach to addressing strong electron-phonon coupling begins with a disentangling unitary transformation that rotates the system into a frame where the electron and phonon wavefunctions are approximately separable into a product state. Following this, the analysis is typically restricted to a low-energy manifold of the phonons degrees of freedom within this rotated frame. This approach yields an effective electron/spin Hamiltonian, with its parameters dressed by the phonon bath~\cite{Takada_2003,Barone_2006,Reja_2011,Reja_2012,Reja_2018,Stolpp_2020,Lu_2023,Islam_2024,Han_2020,Wang_2021,Han_2023,Wang_2020,Han_2024,Ghosh_2017}. Different phases of matter can then be characterized by studying the ground state of this effective Hamiltonian within a pure-state frame.

\begin{figure}[htbp]
\fontsize{6}{10}\selectfont 
\centering
\includegraphics[width=1.0\columnwidth]{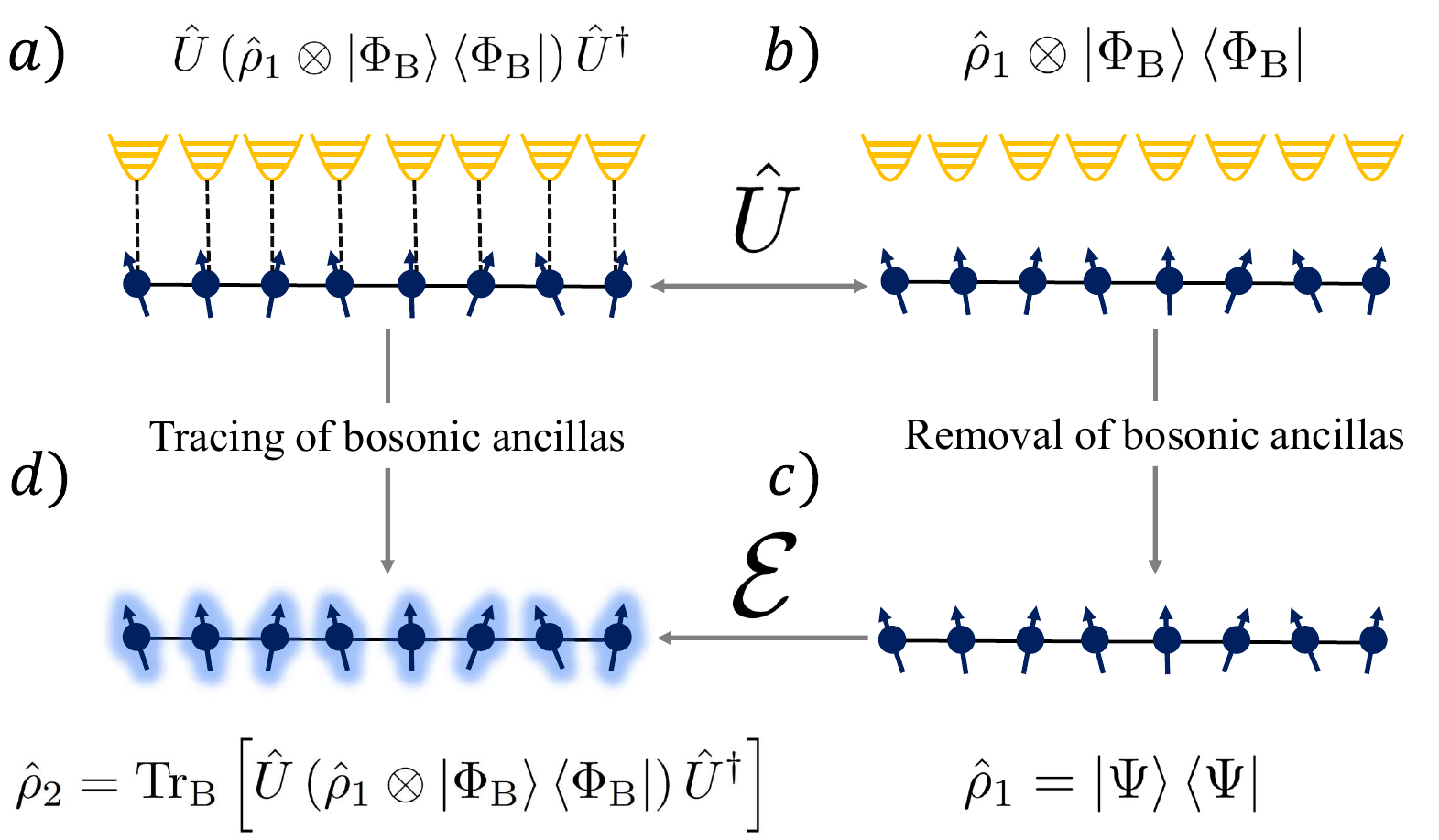}
\caption{Pure- $\hat{\rho}_1$ and mixed- $\hat{\rho}_2$ states in the spin-Holstein-type models. $a)$ Due to the (potentially strong) spin-phonon interaction, the system resides in an 
entangled spin-phonon state. As an ansatz, we assume it can be
written as an entangling transformation with the unitary operator $\hat{U}$ acting on
an arbitrary spin pure-state $\hat{\rho}_1$ and a bosonic pure-state $\ket{\Phi_B}\bra{\Phi_B}$. $b)$ In the pure state approach, we disentangle the spin and bosonic pure-states by $\hat{U}^\dagger$ (polaron type unitary transform). $c)$ Properties of $\hat{\rho}_1=\ket{\Psi}\bra{\Psi}$ can be analyzed in the polaron rotated frame where $\ket{\Psi}$ and $\ket{\Phi_B}$ are approximately unentangled. $d)$ We also analyze $\hat{\rho}_2$, the ``Boson corrupted" mixed state by explicitly tracing out the bosonic degrees of freedom of the spin-phonon coupled state. 
States $a)$ and $b)$ are connected via a unitary transform, while $d)$ can be achieved from $c)$ via a quantum channel $\mathcal{E}$.}
\label{fig:figure 1}
\end{figure}

On a separate note, there has been a recent surge of interest in exploring mixed-state analogues of the concept of many-body phases and their transitions, such as topological order~\cite{Fan_2024,Wang_2024,Bao_2023,Li_2024,Sohal_2024,Ellison_2024,Lee_Moon_2024} and symmetry protected topological phases~\cite{Lake_2023,de_Groot_2022,Zhang_Jian_Hao_2024,Ma_2023,Lee_2024,Roberts_2017,Xue_2024,Guo_Yuchen_2024,Ma_2024_Doubled,McGinley_2020,Deng_2021,Guo_2024}. Indeed, various effects of system-environment interactions naturally require mixed-state description; examples include  decoherence~\cite{Sang_2024,Zou_2023,Lee_2023,Ma_2023_Keldysh,Bombin_2012,Wang_Ting_2024}, dissipation~\cite{Mao_2024,Liu_2024,Min_2024}, measurement~\cite{Su_2024_higher_form,Hauser_2024}, and finite temperature~\cite{Sang_2024,Kimchi_2018,Grusdt_2017,Huang_2014,Viyuela_2014,Rivas_2013,Roberts_2017,McGinley_2020,Bardyn_2018,Liu_2024,Chen_2024_enforced,Guo_2024_new_framework,Lessa_2024_SW,Lu_2020,Weinstein_2024,Molignini_2023,Min_2024,HMB11,Kim_2024}. Furthermore, a novel type of spontaneous symmetry breaking, only present in mixed states, has been studied in Refs.~\cite{Ando_2024,Lessa_2024_SW, Sala_2024,Kuno_2024,Zhang_2024,Su_2024_choi_spin_liquds}. These studies are motivated by the inevitable and often uncontrollable interaction between a many-body system and its environment. For example, a topologically ordered ground state of a local gapped Hamiltonian will not retain its purity under local decoherence. Therefore, an important question arises: Can desirable
properties of topologically ordered states--such as long-range entanglement and exotic excitations--be preserved in more realistic settings?

To define phases and their transitions, it is essential to establish an equivalence relation, that is, to determine when two phases are considered the same~\cite{Coser_2019}. For two pure states in the same phase, described by $\ket{\psi_1}$ and $\ket{\psi_2}$, there must exist a finite-depth local unitary evolution that connects them \cite{Chen_2010}: $\ket{\psi_2} = \hat{U} \ket{\psi_1}$. Analogously, a similar equivalence relation has been proposed for mixed states. For example, it has been argued that for two density operators, $\hat{\rho}_1$ and $\hat{\rho}_2$, to be considered in the same phase, there must exist finite-depth local quantum channels, $\mathcal{E}_{2\leftarrow 1}$ and $\mathcal{E}_{1\leftarrow 2}$, such that $\hat{\rho}_2 = \mathcal{E}_{2\leftarrow 1}[\hat{\rho}_1]$ and $\hat{\rho}_1 = \mathcal{E}_{1\leftarrow 2}[\hat{\rho}_2]$ \cite{HMB11,KR14,CA19,Sang_2024_QMI}. 
We note here that quantum channels are generally noninvertible unlike pure states, where the unitary transformation is always invertible. The concept of this two-way finite-depth channel connectivity between mixed states can be considered as an extension of the pure-state equivalence relation and has led to the exploration of different measures to probe transitions between mixed states. We emphasize, however, that two-way channel connectivity is not the only existing definition of an equivalence relation. In fact, an equivalence relation in the pure-state sense could still be useful when representing a mixed-state density matrix as a pure-state in a doubled Hilbert space \cite{Sala_2024,Zhang_2024}.

With the proposed equivalence relations for mixed states, the next task is to develop diagnostic tools that can probe transitions between different phases. Several candidates have been suggested, each based on slightly different notions of mixed-state phases depending on the equivalence relation upon which they are based. In particular, the concept of a finite Markov length (denoted by $\xi$) has been proposed as a possible criterion for inferring mixed-state phase transitions \cite{Sang_2024_QMI,Zhang_2024,Negari_2024}. This measure could serve as a defining property, analogous to the finite energy gap between a ground state and excited states in gapped local Hamiltonians. The measure is defined as a length scale, $\xi$, at which the conditional mutual information (CMI) defined on a tripartite system (see Fig.~\ref{fig:figure 3}) decays exponentially: CMI~$\sim \exp(-r/\xi)$. In this work, we focus on two variants of the CMI: the von Neumann CMI and the R\'enyi-2 CMI. 

The von Neumann CMI is based on the two-way channel connectivity equivalence relation, utilizing the connection between a finite von Neumann Markov length $\xi_{\text{vN}}$ and a recovery map \cite{Petz_1986} that can reverse the channel mapping from one density matrix to another. For example, the critical behavior of the Toric code under dephasing noise has been explored by identifying a finite dephasing error rate, diagnosed by the divergence of $\xi_{\text{vN}}$ \cite{Sang_2024_QMI}. 

The R\'enyi-2 CMI is based on the quasi-local unitary connectivity equivalence relation in the doubled Hilbert space, utilizing the Choi-Jamio\l kowski isomorphism \cite{MDC75,AJ72}. The divergence of the R\'enyi-2 Markov length $\xi_{2}$ indicates a phase transition in the doubled Hilbert space, which naturally connects to the divergent correlation length in the R\'enyi-2 correlators in the original Hilbert space. It has been applied to examine transitions in the Toric code with various types of decoherence \cite{Zhang_2024}.

Both von Neumann and Rényi-2 CMIs have been proposed as promising candidates for diagnostic tools for mixed-state phase transitions. Although similar in construction, there is a subtle difference between the two, and the choice of which to utilize is still an active topic of discussion. Furthermore, this subtlety becomes even harder to resolve with the enriched types of symmetries present in mixed states. 

Unlike pure states, mixed states possess two types of symmetries: \textit{weak} and \textit{strong}~\cite{Lessa_2024,Wang_2024_anomaly,Hsin_2023}. In the case of weak symmetry, a symmetry operator $\hat{W}$—typically unitary—acts on the density matrix as $\hat{W} \hat{\rho} \hat{W}^\dagger = \hat{\rho}$. This means that when $\hat{\rho}$ is expressed in its spectral decomposition, $\hat{\rho} = \sum_n p_n \ket{\psi_n} \bra{\psi_n}$, the operator $\hat{W}$ affects each eigenstate $\ket{\psi_n}$ by introducing a phase factor: $\hat{W} \ket{\psi_n} = e^{i \phi_n} \ket{\psi_n}$. On the other hand, strong symmetry is defined by a symmetry operator that acts directly on the density matrix as $\hat{W} \hat{\rho} = e^{i \phi} \hat{\rho}$, which implies that $\hat{W}$ introduces a global phase across all eigenstates, such that $\hat{W} \ket{\psi_n} = e^{i \phi} \ket{\psi_n}$ for every $\ket{\psi_n}$. With two types of symmetries, it has been suggested that a novel type of spontaneous symmetry breaking could occur in mixed state, where a strong symmetry is spontaneously broken to a weak one. This strong-to-weak spontaneous symmetry breaking (SW-SSB) has been argued as a universal property of mixed-state phase transitions  \cite{Lessa_2024_SW,Ma_2024,Guo_Yuchen_2024,Ma_2024_Doubled,Sohal_2024,Sala_2024,Xu_2024,Huang_2024_hydro,Gu_2024,Moharramipour_2024,Su_2024_choi_spin_liquds,Kuno_2024,Shah_2024,Lu_2024,Ando_2024,Zhang_2024}. An important open question is how the divergent $\xi$ from both von Neumann and R\'enyi-2 CMIs can be precisely related to SW-SSB. Although the connection between the divergent Markov length and SW-SSB will not be discussed in this work, it would be an interesting future direction.

With the advent of novel measures capable of probing mixed-state phase transitions, a natural question arises: Could these measures offer a more comprehensive understanding of phase transitions in coupled electron/spin-phonon systems? For instance, when characterizing the ground-state phase of an electron/spin system strongly coupled to environmental phonon modes, the standard approach 
is to transform the entangled electron/spin-phonon state of the system to the frame
where the state can be approximated by a product state
of the electron/spin and boson wavefunctions, $\ket{\Psi}$ and $\ket{\Phi_B}$.  This corresponds to undoing the entangling unitary transformation $\hat{U}$ via $\hat{U}^\dagger=\hat{U}_p$, where $\hat{U}_p$ is the polaron disentangling unitary (see Fig.~\ref{fig:figure 1} panels (a)-(b)). The different phases can be characterized by solely examining the electron/spin pure ground state $\ket{\Psi}$, by projecting out the phonon part $\ket{\Phi_B}$; see Fig.~\ref{fig:figure 1}(c).
Alternatively, by treating the bath phonon modes as ancillary degrees of freedom, we can obtain a mixed state for the system by explicitly tracing out the bosons; see Fig.~\ref{fig:figure 1}(d). By characterizing distinct phases of matter described by this latter ``phonon-corrupted" mixed state through different versions of the CMI, we aim to extend the notions of mixed-state-phases and their classification to the realm of electron/spin-phonon systems.   

Specifically, we consider the two-dimensional (2D) cluster state on a Lieb lattice, locally coupled to harmonic degrees of freedom via an arbitrary coupling strength; see Fig.~\ref{fig:figure 2}b). The 2D cluster state is a prototypical example of a symmetry-protected topological (SPT) phase that acts as a resource state for measurement-based quantum computation \cite{Gross_2007,Browne_2003,Briegel_2001,Raussendorf_2001,Wei_2003,Raussendorf_2003,Barret_2004,Grover_1997,Monroe_1997}. 
The topological phase transition to a trivial phase, upon exceeding a finite decoherence strength has already been detected using distinct diagnostic methods \cite{Guo_2024}. However, different notions of mixed state phases probed by different versions of Markov length analysis in the electron/spin-phonon setting have yet to be explored. Unlike this previous study that only effectively included environmental effects through quantum channels, we begin here with a microscopic model for the coupled electron/spin-phonon model. Considering the cluster Hamiltonian with local phonons, we first show that the pure-state framework fails to detect the anticipated transition to a topologically trivial phase at strong spin-phonon coupling. In contrast, in the mixed-state picture we employ different diagnostics for mixed-state phase transitions, the von Neumann CMI and the Rényi-2 CMI, detect signatures of mixed-state phase transitions, and explore different notions of mixed-state phases.

The rest of the paper is organized as follows. In Sec.~\ref{sec: Model Hamiltonian}, we introduce a concrete model of a system strongly coupled to its environment, which can be described as a spin-Holstein model, namely, the 2D cluster Hamiltonian locally coupled to 
bosonic modes. In Sec.~\ref{sec: Pure-state approach}, we demonstrate a standard method to obtain an effective spin-only Hamiltonian by performing a unitary decoupling transformation, followed by a truncation to the low-energy manifold of the environmental degrees of freedom. We then study the phase of the corresponding ground state wavefunction of the effective Hamiltonian as a function of the system-environment coupling strength. In Sec.~\ref{sec: Mixed-state approach}, we consider the original frame of reference, where the system strongly couples to the environment, and explicitly trace out the environmental degrees of freedom. This results in a phonon-decohered mixed-state density matrix, whose phase is analyzed by using two types of CMI via analytical method for the von Neumann and numerical method for R\'enyi-2, respectively. We give a summary and conclude in Sec.~\ref{sec: Discussion and conclusion}. Technical details are delegated to the Appendices.


\section{spin-holstein cluster Hamiltonian}
\label{sec: Model Hamiltonian}

In this section, we introduce a model Hamiltonian for a coupled electron-phonon system. Specifically, we introduce the spin-Holstein cluster model on the 2D Lieb lattice, where each lattice site is coupled to a single harmonic mode with arbitrary coupling strength, and
describe its ground state. 

The total Hamiltonian for this system is given by $\hat{H}=\hat{H}_S+\hat{H}_I+\hat{H}_B$, where the spin system, spin-bath interaction, and bath Hamiltonians are given as follows:
\begin{equation}
\label{eq: total hamiltonian}
\begin{aligned}
    \hat{H}_S =& -\sum_v\hat{X}_v \hat{Z}_{e_1} \hat{Z}_{e_2} \hat{Z}_{e_3} \hat{Z}_{e_4}-\sum_e\hat{X}_e \hat{Z}_{v_1} \hat{Z}_{v_2},\\
    \hat{H}_I =& g\sum_v\hat{\tau}^x_v\left(\hat{b}^\dagger_v+\hat{b}_v\right)+g\sum_e\hat{\sigma}^x_e\left(\hat{b}^\dagger_e+\hat{b}_e\right),\\
    \hat{H}_B = &\omega \sum_v\hat{b}^\dagger_v\hat{b}_v+\omega\sum_e\hat{b}^\dagger_e\hat{b}_e.
    \end{aligned}
\end{equation}
In what follows we therefore refer to interactions of the lattice with the phonon modes as spin-phonon or spin-boson couplings, rather than using the language of ``electron-phonon coupling".

Let us now describe each contribution in Eq.~\eqref{eq: total hamiltonian} in more detail.
\begin{figure}[b]
\fontsize{6}{10}\selectfont 
\centering
\includegraphics[width=1.0\columnwidth]{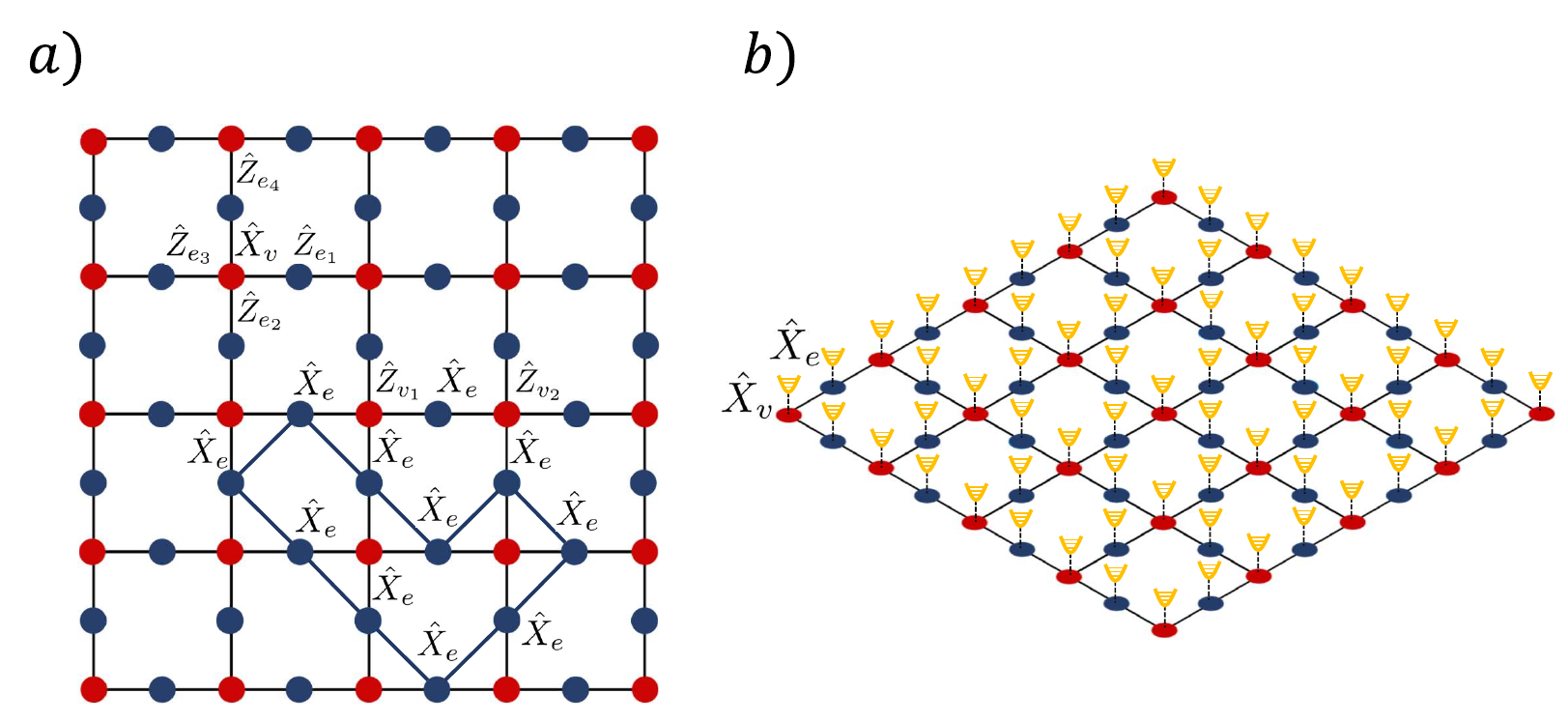}
\caption{a) 2D cluster model on the Lieb lattice. The vertex term $\hat{A}_v$ (top-left) and the edge term $\hat{B}_e$ (middle) are shown. The blue loop indicates an example of $\mathds{Z}^{(1)}_2$ symmetry. b) The 2D cluster model with local interactions to bosonic modes via $\hat{X}_i$ at each lattice site.}
\label{fig:figure 2}
\end{figure}

i) $\hat{H}_S$ is a frustration-free stabilizer Hamiltonian on the edge-decorated square lattice, consisting of two nonequivalent sublattices (see Fig.~\ref{fig:figure 2}a)). Sublattice $A$ is located at the vertices (red) of an $N\times N$ square lattice, while sublattice $B$ is placed at the edges (blue) of the square lattice. Both sublattices, $A$ and $B$, are occupied by spin-$1/2$ particles, denoted by Pauli operators $\{\hat{X}_i,\hat{Y}_i,\hat{Z}_i\}$ with site index denoted as subscript $i\in\{e,v\}$. The vertex and edge terms of the spin system Hamiltonian are denoted $\hat{A}_v = \hat{X}_v \hat{Z}_{e_1} \hat{Z}_{e_2} \hat{Z}_{e_3} \hat{Z}_{e_4}$ and $\hat{B}_e = \hat{X}_e \hat{Z}_{v_1} \hat{Z}_{v_2}$, respectively. These operators satisfy the conditions $[\hat{A}_v, \hat{B}_e] = 0$ and $\hat{A}^2_v =\hat{B}^2_e = 1$ for any vertex $v$ and edge $e$.

Under periodic boundary conditions (PBCs), the unique ground state of $\hat{H}_S$ is characterized by setting $\hat{A}_v = \hat{B}_e = 1$ for all $v$ and $e$, leading to the ground state projector $\hat{\rho}^\text{(PBC)}_\text{SPT} = \prod_{v\in A} \frac{1 + \hat{A}_v}{2} \prod_{e\in B} \frac{1 + \hat{B}_e}{2}$. The unique ground state given by the projector is protected by the $\mathds{Z}^{(0)}_2 \times \mathds{Z}^{(1)}_2$ symmetry. The global zero-form symmetry $\mathds{Z}^{(0)}_2$ is defined on the vertex sublattice $A$, and is generated by $\prod_{v \in A} \hat{X}_v$. The one-form symmetry $\mathds{Z}^{(1)}_2$ is defined in loops in the edge sublattice $B$, generated by $\prod_{e \in \text{loop on } B} \hat{X}_e$ (see Fig.~\ref{fig:figure 2}a)), resulting in an exponentially large number of conserved charges in the thermodynamic limit. For an open boundary condition, the cluster state has boundary degrees of freedom that are associated with ground-state degeneracy.

ii) $\hat{H}_I$ describes the coupling between the spin system and local bosonic modes; see Fig.~\ref{fig:figure 2}(b). For simplicity, we assume the spin-boson coupling strength $g$ to be the same for all sites. Each site $i\in\{v,e\}$ is coupled to a bosonic displacement operator $\hat{b}^\dagger_i+\hat{b}_i$ where $\hat{b}^\dagger_i$~($\hat{b}_i$) is the creation~(annihilation) operator for the environment modes satisfying $[\hat{b}_i,\hat{b}^\dagger_j]=\delta_{ij}$. 

iii) $\hat{H}_B$ describes the harmonic oscillator modes on each site. For simplicity, we assume that all modes have an identical frequency, $\omega$. 

\section{Pure-state approach}
\label{sec: Pure-state approach}
In this section, we review the conventional approach taken by numerous previous studies to characterize the different ground-state phases in strongly correlated electron systems subject to electron-phonon interactions (see, e.g., Ref.~\cite{Hohenadler_2007}).

When the coupling strength between the system and the environment is arbitrary, the standard approach for studying the system's ground state involves applying a unitary transformation to rotate the entangled spin-phonon state to a frame where the spin system and the environment phonons are largely decoupled. This is followed by a projection onto the low-energy manifold of the environmental phonon degrees of freedom. The result of this transformation and truncation is a spin-system Hamiltonian, where the effects of strong spin-phonon coupling are reflected in the modification of the spin system's parameters, which now depend on the environmental-phonon parameters.

Specifically, we define an effective system Hamiltonian as follows:
\begin{equation}
\hat{H}^\text{eff}_S=\bra{0}\hat{U}_p\hat{H}\hat{U}^\dagger_p\ket{0}.
\label{eq:HSeff}
\end{equation}
Here, $\ket{0}=\bigotimes_i\ket{0_i}$ is a product state of ground states of harmonic oscillators, $\hat{b}_i\ket{0_i}=0$ and $\hat{U}_p=\prod_i\hat{U}_i$ is a product of (commuting) local polaron transformations, $\hat{U}_i=e^{\frac{g}{\omega}\hat{X}_i(\hat{b}^\dagger_i-\hat{b}_i)}$. After the unitary transformation, the total Hamiltonian reads
\begin{equation}
  \begin{aligned}  \hat{U}_p\hat{H}\hat{U}^\dagger_p=-&\sum_v\hat{X}_v\prod^4_{j=1}\left(\hat{\mathcal{C}}_{e_j}\hat{Z}_{e_j}-\hat{\mathcal{S}}_{e_j}\hat{Y}_{e_j}\right)\\
  -&\sum_e\hat{X}_e\prod^2_{j=1}\left(\hat{\mathcal{C}}_{v_j}\hat{Z}_{v_j}-\hat{\mathcal{S}}_{v_j}\hat{Y}_{v_j}\right)\\
  +&\omega\sum_i\hat{b}^\dagger_i\hat{b}_i+\text{const},
  \end{aligned}
\end{equation}
where $\hat{\mathcal{C}}_{i}=\cos(\frac{2ig}{\omega}(\hat{b}^\dagger_i-\hat{b}_i))$ and $\hat{\mathcal{S}}_i=\sin(\frac{2ig}{\omega}(\hat{b}^\dagger_i-\hat{b}_i))$. The projection onto the harmonic modes ground state, assuming high-frequency bath modes, allows us to generate the effective spin-system Hamiltonian,
\begin{equation}
\label{eq: effh}
    \hat{H}^\text{eff}_S=-\kappa_v\sum_v\hat{A}_v-\kappa_e\sum_e\hat{B}_e
\end{equation}
where $\kappa_v=\exp(-8g^2/\omega^2)$ and $\kappa_e=\exp(-4g^2/\omega^2)$,
see Appendix~\ref{sec: Derivation for the effective system Hamiltonian} for a detailed derivation.
Since both $\kappa_v$ and $\kappa_e$ are strictly positive, the ground state of $\hat{H}^\text{eff}_S$ is still given by setting $\hat{A}_v=\hat{B}_e=1$, regardless of the coupling strength $g$. 
We denote this ground state by $\ket{\tilde{\Psi}_\text{GS}}$, and we introduce the associated pure state density matrix,
\bea
\hat \rho_1  =  \ket{\tilde{\Psi}_\text{GS}}  \bra{\tilde{\Psi}_\text{GS}}.
\eea
The ground state wave function for the entire system in the polaron frame (denoted with tilde), $\ket{\tilde{\Psi}_\text{tot}}$,  can therefore be well approximated by a product state of the 2D cluster state, $\ket{\tilde{\Psi}_\text{GS}}$ and the vacuum states of the bath bosonic modes, 
\begin{equation}
\ket{\tilde{\Psi}_\text{tot}}=\ket{\tilde{\Psi}_\text{GS}}\otimes\ket{0}.
\end{equation}

It is useful to note that an optimal choice for $\ket{\tilde{\Psi}_\text{tot}}$ can be attempted by a variational process. This is done by introducing two variational parameters, $\lambda$ and $\alpha$ for the polaron transformation and for the low-energy manifold truncation, respectively. That is, we write the polaron unitary with the variational parameter $\lambda$, $\hat{U}_i(\lambda)=e^{\frac{\lambda}{\omega}\hat{X}_i(\hat{b}^\dagger_i-\hat{b}_i)}$, and the bosonic wavefunction as a coherent state, $\hat{b}_i\ket{\alpha_i}=\alpha\ket{\alpha_i}$. The optimal values for $\lambda$ and $\alpha=x+ip$ can be found by minimizing the energy $E(\lambda,x,p)=(\bra{\alpha}\otimes\bra{\tilde{\Psi}_\text{GS}})\hat{U}_\lambda \hat{H}\hat{U}^\dagger_\lambda(\ket{\tilde{\Psi}_\text{GS}}\otimes\ket{\alpha})$. Here, $\ket{\alpha}=\otimes_i\ket{\alpha_i}$ and $\hat{U}^\dagger_\lambda=\prod_i\hat{U}_i(\lambda)$. 
However, as we show in Appendix~\ref{sec: variational process}, using this approach the optimal parameters turn out to be $\alpha=0$. The optimal choice for $\lambda$ will not affect the overall structure of the effective spin-Hamiltonian as described in Eq.~\eqref{eq: effh}, nor its ground state, as it simply replaces $g$ with $\lambda$ for $\kappa_{v/e}$.

The approach presented here can be understood as a canonical transformation from a frame where the ground state wavefunction is represented by highly entangled spins and phonon components, as depicted in Fig.~\ref{fig:figure 1}~$a)$, to a frame where the ground state can be approximately 
described by a product state of the two components, as shown in Fig.~\ref{fig:figure 1}~$b)$. However, the finding that the cluster state remains the unique stable ground state, regardless of the spin-boson coupling strength (see discussion below Eq.~(\ref{eq: effh})), is at odds with the physical intuition that the SPT order of the 2D cluster state should be lost when the coupling strength exceeds a certain critical value. 

\section{Mixed-state approach}
\label{sec: Mixed-state approach}

In this section, we follow the mixed-state approach to characterizing the ground-state phase of the 2D cluster Hamiltonian that is locally coupled to bosonic modes, as described in Eq.~\eqref{eq: total hamiltonian}. 
Since $\ket{\tilde{\Psi}_\text{tot}}=\ket{\tilde{\Psi}_\text{GS}}\otimes\ket{0}$ is the ground-state wavefunction in the polaron-rotated frame, if we go back to the original frame, the ground-state wave function is given as $\ket{\Psi_\text{tot}}=\hat{U}^\dagger_p\ket{\tilde{\Psi}_\text{tot}}$, which will entangle the spin and bosonic wavefunctions. When we trace out the bosonic degrees of freedom, we end up with a (spin-only) mixed state, 
\bea
\hat{\rho}_2 &=& \text{Tr}_\text{B}(\ket{\Psi_\text{tot}}\bra{\Psi_\text{tot}})
\nonumber\\
&= &
\text{Tr}_\text{B}\left[ \hat U_p^{\dagger} 
\left(\hat \rho_1 \otimes\ket {0}\bra{0}\right)\hat U_p\right].
\eea
As we show in Appendix~\ref{sec: Details on the mixed state derivation}, this mixed state can be written as
\begin{equation}
\label{eq: mixed state rho2}
\hat{\rho}_2
=\mathcal{E}_{2\leftarrow1}[\hat{\rho}_1]=\prod_i\left[(1-p)\hat{\rho}_1+p\hat{X}_i\hat{\rho}_1\hat{X}_i\right].
\end{equation}
Here, 
\bea p=\frac{1}{2}[1-e^{-2(g/\omega)^2}],
\label{eq:pgw}
\eea
is a parameter that describes the bath-induced bit-flip (error) rate. This equation reflects the connection between the microscopic spin-phonon coupling and the resulting error channel corrupting the cluster state. Eq.~\eqref{eq: mixed state rho2} describes a quantum channel $\mathcal{E}_{2\leftarrow 1}$ that takes a pure state $\hat{\rho}_1$ with SPT order, and produces a mixed state $\hat{\rho}_2$ through a single-qubit flip error with rate $p$. Our goal is to determine whether there exists a critical value of $p$ (or equivalently, the coupling strength $g/\omega$) that signals a phase transition from an SPT phase to a topologically trivial phase while using a mixed-state framework.

A previous study \cite{Guo_2024} directly modeled bit flip errors using the channel approach described in Eq.~(\ref{eq: mixed state rho2}), without explicitly deriving it from the underlying spin-phonon interaction model. It demonstrated that there exists a critical value of $p=p_c\sim 0.1782$ (equivalently $g/\omega\sim 0.4694$), where the system undergoes a topological phase transition from the $\mathds{Z}^{(0)}_2\times \mathds{Z}^{(1)}_2$ SPT phase to the trivial phase. This result was obtained by mapping the 2D cluster state under local bit-flip error to a classical 2D Ising model and then characterizing the phase transition using three diagnostic measures (relative entropy, strange correlation function, and multipartite negativity), yielding consistent results.  

\begin{figure}[b]
\fontsize{6}{10}\selectfont 
\centering
\includegraphics[width=0.8\columnwidth]{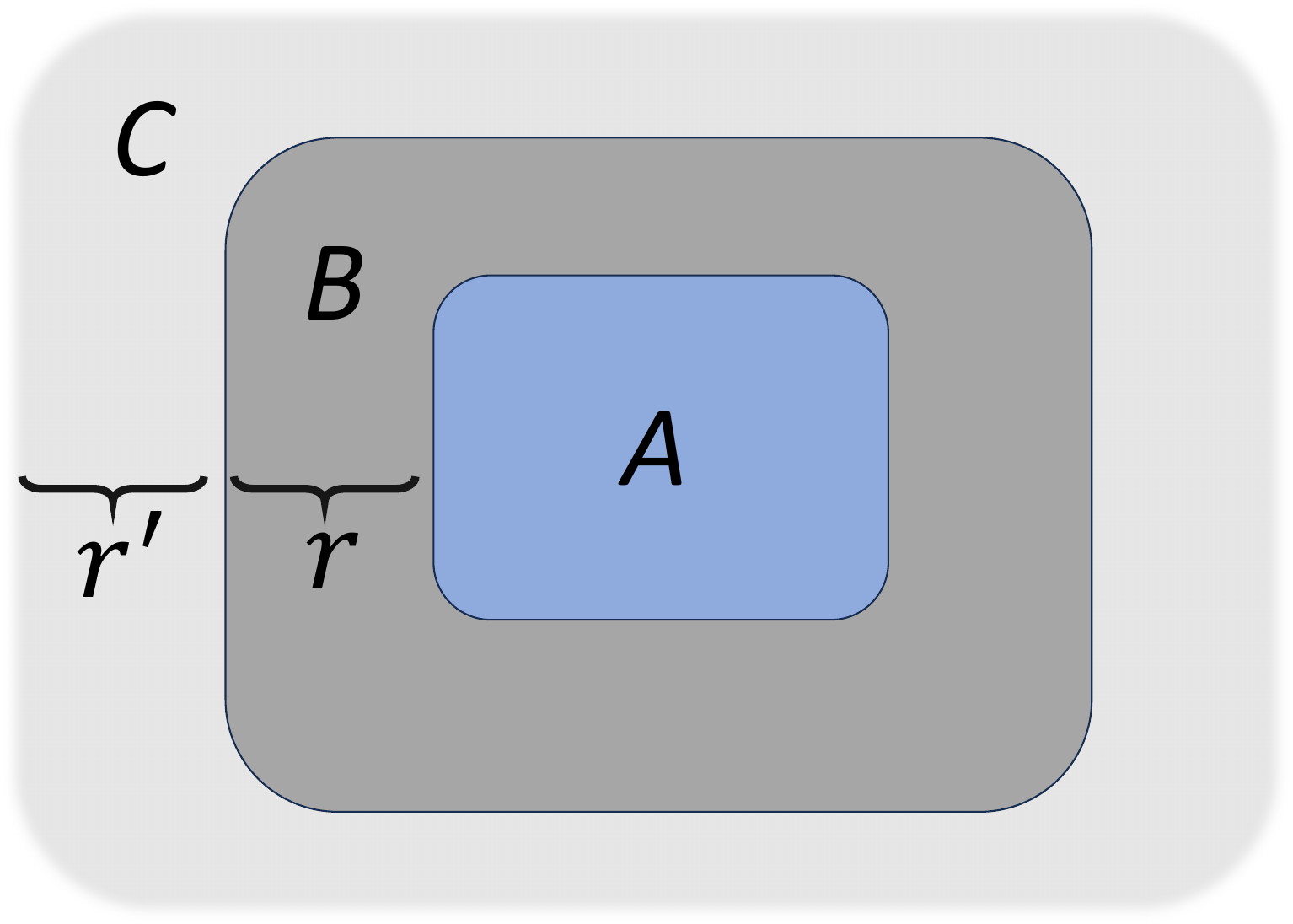}
\caption{Tripartition of regions where the CMI is defined. The radius of region $A$ is fixed while $r$ and $r'$, which define regions B and C, respectively, are taken to be arbitrary but large.}
\label{fig:figure 3}
\end{figure}

In the present work, we employ a recently proposed marker for mixed-state phase transitions, capable of distinguishing different phases based on the scaling behavior of the quantum conditional mutual information (CMI) defined on a tripartite geometry; see Fig.~\ref{fig:figure 3}. We employ two types of CMI. 
First, in Sect.~\ref{sec: von neumann} we use the von Neumann CMI and present an analytical argument supporting the existence of a phase transition at a certain value of the bit-flip rate $p$. In Sec.~ \ref{sec: renyi-2},
we study the R\'enyi-2 CMI through numerical simulations and validate analytical results by similarly predicting the existence of a single phase transition.

i) The von Neumann CMI is defined as 
\begin{equation}\label{vnI}
    I_{\text{vN}}(A:C|B)=[S(AB)-S(B)]-[S(ABC)-S(BC)]
\end{equation}
where $S(M)$ is the von Neumann entanglement entropy between region $M$ and its complement region $\overline{M}$. CMI on the geometry in Fig.~\ref{fig:figure 3} essentially measures how nonlocal $A$'s correlation is between its complement. The expected scaling behaviors of $I_{\text{vN}}(A:C|B)$
as a function of $r$, the size of the region $B$, are distinct: exponential decay, $I_{\text{vN}}(A:C|B)\sim \exp(-r/\xi_{\text{vN}})$, away from critical points, and power-law decay near the critical points. These distinct scalings can serve to distinguish between different mixed-state phases, the SPT and the trivial phases in the current study. In particular, it has been shown that the finiteness of the von Neumann Markov length $\xi_{\text{vN}}$ throughout the time evolution generated by a local Lindbladian implies the two-way connectivity via (quasi)local channels $\mathcal{E}_{2\leftarrow 1}$ and $\mathcal{E}_{1\leftarrow 2}$ \cite{Sang_2024_QMI}. 

ii) The R\'enyi-2 CMI is defined as \cite{Zhang_2024}
\begin{equation}
\begin{aligned}
     I_2(A:C|B)=&[S^{(2)}(AB)-S^{(2)}(B)]\\
     -&[S^{(2)}(ABC)-S^{(2)}(BC)],
     \end{aligned}
\end{equation}
which has the same structure as the von Newmann CMI, $I_{\text{vN}}(A:C|B)$, but each component is replaced by the R\'enyi-2 entropy, $S^{(2)}(M)=-\ln[\Tr(\hat{\rho}^2_M)]$. Here, $\hat{\rho}_M$ is a reduced density matrix of region $M$. The R\'enyi-2 Markov length $\xi_{2}$ can also be defined in the same manner as above, namely, $I_2(A:C|B)\sim\exp(-r/\xi_2)$ away from critical points. 
The divergence of $\xi_2$ has a simple interpretation as the divergent correlation length in the R 'enyi-2 correlator, i.e., the correlation functions in the double Hilbert space for a mixed state via the Choi-Jamio\l kowski isomorphism. 

As mentioned earlier, both the von Neumann and Rényi-2 versions of the CMI can serve as diagnostics for mixed-state phase transitions, yet they reflect distinct perspectives. Moreover, these diagnostics are not the only possibility; other approaches, such as fidelity correlators, also exist \cite{Hu_2024,Kim_2024}. Each of these diagnostics has its respective strength and limitation. For example, for the Markov length $\xi_{\text{vN}}$ captured by the von Neumann CMI, its finiteness implies a two-way channel connectivity, while the converse does not necessarily hold~\cite{Zhang_2024}. In addition, determining the exact critical value using the divergent $\xi_{\text{vN}}$ is computationally challenging. Nevertheless, the von Neumann CMI allows for a clear information-theoretic interpretation. In contrast, identifying the divergent Markov length $\xi_{2}$ defined in terms of the R\'enyi-2 CMI is computationally more tractable than in the von Neumann CMI. The divergent $\xi_{2}$ also has a one-to-one correspondence to the divergent correlation length in the Choi state in the doubled Hilbert space, and it allows us to distinguish distinct mixed phases in a similar manner as in pure-state phases. However, its role as a definitive marker for the recoverability or channel connectivity of an error-decohered state currently remains unclear.

\subsection{Analytical results on von Neumann CMI}
\label{sec: von neumann}

Here, we identify the von Neumann entropy as the  temperature-scaled free energy of a finite-size random-bond
Ising model.  Based on this picture for the von Neumann CMI, we then discuss mixed-state phases of the system from a perspective of the two-way channel connectivity. 

We aim to calculate the von Neumann entropy of the reduced density matrix $\hat{\rho}_M$ of a decohered 2D cluster state, corresponding to an arbitrary choice of region $M$, the latter is defined by 
\begin{equation}\label{eq: noisey rdm}
    \hat{\rho}_M = \prod_{i\in M}\mathcal{N}^X_i\left[\prod_{v\in M}\frac{1+\hat{A}_v}{2}\prod_{e\in M}\frac{1+\hat{B}_e}{2}\right],
\end{equation}
Here, $\mathcal{N}^X_i[\bullet]$ is a local bit-flip error channel defined as $\mathcal{N}^X_i[\bullet]=(1-p)\bullet+p\hat{X}_i\bullet \hat{X}_i$, see
Eq.~(\ref{eq: mixed state rho2}). The $n$th R\'enyi entropy is defined by
\begin{equation}
    S^{(n)}(M) = -\frac{1}{n-1}\ln\Tr(\hat{\rho}^n_M).
\end{equation}
The von Neumann entropy of region $M$ (see Fig.~\ref{fig:figure 3}) can be regarded as the replica limit of $n\rightarrow 1$, 
\begin{equation}
    S(M) = -\frac{\partial}{\partial n}\Tr (\hat{\rho}^n_M)\big|_{n=1},
\end{equation}
where $n$ is an integer greater than 1. 

We recall that $\hat{\rho}_M$ can be decomposed into sums of domain-wall configurations and gauge configurations \cite{Guo_2024}. Specifically, the corresponding (classical) statistical mechanical model for the $n$ replica consists of a sum of the $(n-1)$-flavor Ising Hamiltonian and the $(n-1)$-flavor Ising gauge Hamiltonian:
    \begin{eqnarray}
        {H}_\text{Ising}^{(n)} &=& -J_n\sum_{\langle ij\rangle}\left(\sum_{\alpha=1}^{n-1}s_i^{(\alpha)}s_j^{(\alpha)}+\prod_{\alpha=1}^{n-1}s_i^{(\alpha)}s_j^{(\alpha)}\right),\label{isingH}\\
        {H}_\text{Gauge}^{(n)}&=&-U_n\sum_{\square}\left(\sum_{\alpha=1}^{n-1}\phi_{\square}^{(\alpha)}+\prod_{\alpha=1}^{n-1}\phi_{\square}^{(\alpha)}\right),\label{gaugeH}
    \end{eqnarray}
where $s^{(\alpha)}_i\in\{-1,1\}$, $\phi^{(\alpha)}_{\square}=\prod_{i\in\square}s^{(\alpha)}_i$, and $J_n$ and $U_n$ are coupling strengths determined by the error parameter $p$. The $n$th R\'enyi entropy of region $M$ can then be obtained as the free energy of the above two Hamiltonians,
\begin{equation}
    S^{(n)}(M)=\frac{1}{1-n}\left(F^{(n)}_\text{Ising}(M)+F^{(n)}_\text{Gauge}(M)\right),
\end{equation}
where we define $F^{(n)}_\text{Ising}=-\ln Z^{(n)}_\text{Ising}$ and $F^{(n)}_\text{Gauge}=-\ln Z^{(n)}_\text{Gauge}$ with $Z^{(n)}_\text{Ising}$ and $Z^{(n)}_\text{Gauge}$ being the partition functions of Eqs.~\eqref{isingH} and \eqref{gaugeH}, respectively.  
A non-analytic behavior can be expected from the Ising theory at the critical point, while the Ising Gauge part should not have a phase transition.  

Since our focus is on a phase transition, we focus on the Ising part and attempt to determine the replica limit of $n\rightarrow 1$ of $S^{(n)}(M)$. We note that the Ising model has both high- and low-temperature expansions that are dual to each other. To compute the desired replica limit, we employ the high temperature expansion, leading to 
\begin{equation}
    \hat{\rho}_M \simeq \sum_l P(l)\hat{X}^l\hat{\rho}_0\hat{X}^l,
\end{equation}
where $\hat{\rho}_0=\prod_{v\in M}\frac{1+\hat{A}_v}{2}\prod_{e\in M}\frac{1+\hat{B}_e}{2}$ and $P(l)=p^{|l|}(1-p)^{V_M-|l|}$ is the probability of a string configuration; $l$ is a configuration with one or more open strings in region $M$, $|l|$ is the length of $l$, and $V_M$ is the total number of vortex qubits in region $M$. 
Here, we use $\hat{X}^l$ as a shorthand notation for $\prod_{i\in l}\hat{X}_i$.
The $n$th R\'enyi entropy can be approximated by
\begin{equation}
    \Tr(\hat{\rho}^n_M)\simeq \sum_{l^{(\alpha)}}P(l^{(\alpha)})\Tr\left(\prod^n_{\alpha=1}\hat{X}^{l^{(\alpha)}}\hat{\rho}_0\hat{X}^{l^{(\alpha)}}\right),
\end{equation}
where $\alpha$ represents the replica index.
Note that
\begin{equation}
    \Tr\left(\prod^n_{\alpha=1}\hat{X}^{l^{(\alpha)}} \hat{\rho}_0\hat{X}^{l^{(\alpha)}} \right) = \prod^{n-1}_{\alpha=1}\Tr\left(\hat{X}^{l^{(\alpha)}} \hat{X}^{l^{(\alpha+1)}}\hat{\rho}_0\right)
\end{equation}
and $\Tr\left(\hat{X}^{l^{(\alpha)}} \hat{X}^{l^{(\alpha+1)}}\hat{\rho}_0\right)=1$ if and only if $l^{(\alpha)}\cap l^{(\alpha+1)}$ forms a closed loop; otherwise, the trace is zero. The trace is nonzero if all the strings $l^{(\alpha)}$ for the replicas $\alpha>1$ differ from $l^{(1)}$ up to a closed loop. 
We can regard the difference between $l^{(\alpha)}$ with $\alpha>1$ and $l^{(1)}$ as the boundary of the domain wall $C^{(\alpha)}$. Hence, $\Tr(\hat{\rho}^n_M)$ can be expressed as
\begin{equation}
    \Tr\left(\hat{\rho}^n_M\right)=\sum_{l^{(1)}}P(l^{(1)})\sum_{C^{(\alpha)}}\prod^n_{\alpha=2}P(l^{(1)}+C^{(\alpha)}).
\end{equation}
This can be understood as the $n-1$ copies of the random-bond Ising model (RBIM),
\begin{eqnarray}
    Z^{(n)}_{\text{Ising}}(M)& \sim& \sum_{l^{(1)}}P(l^{(1)})\sum_{\langle ij\rangle} \exp\left(-J_{ij}\sum^n_{\alpha=2}s^{(\alpha)}_is^{(\alpha)}_j\right)\nonumber\\
   & =&Z^{n-1}_\text{RBIM},
\end{eqnarray}
where $J_{ij}=-\frac{1}{2}\ln\frac{p}{1-p}$ if the link between $i$ and $j$ is on the defect line, otherwise $J_{ij}=-J$, and $Z_\text{RBIM}$ is the partition function of the RBIM resulting from the sum over all the defect configurations. Accordingly, taking the limit of $n\to 1$ gives
\begin{equation}
     S(M) = F_\text{RBIM}(M),
 \end{equation}
 where $F_\text{RBIM}=-\ln Z_{\text{RBIM}}$.

We can now summarize our observations regarding the existence of a phase transition.
The von Neumann CMI, $I_{\text{vN}}(A:C|B)$, as expressed in Eq.~\eqref{vnI} can be written as the RBIM free-energy difference from the above analysis, the latter is expected to exhibit a nonanalytic behavior at the RBIM transition point $p_{\text{vN}}\simeq 0.11$ \cite{HA01}; a similar transition has been numerically checked by the von Neumann CMI of the decohered Toric code~\cite{Sang_2024_QMI}.
 This transition, signaled by the divergent von Neumann Markov length, $\xi_{\text{vN}}$, corresponds to a mixed-state transition characterized in terms of the two-way channel connectivity.
Thus, the von Neumann CMI indicates a phase transition at $p_{\text{vN}}\simeq 0.11$ or equivalent to $g/\omega\sim 0.352$, when translating this point to the microscopic spin-phonon parameters using Eq.~(\ref{eq:pgw}).

\begin{figure*}
\includegraphics[width=.96\textwidth]{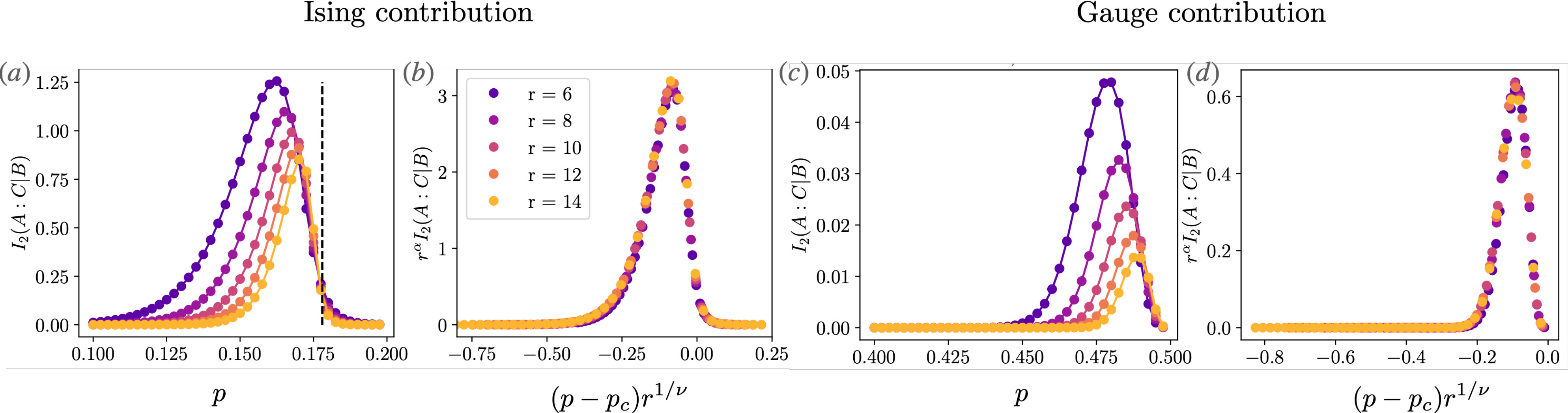}
\caption{R\'enyi-2 CMI in a tripartite system. Contributions from the Ising Hamiltonian and Gauge Hamiltonian are shown in separate plots. Region A is taken to be a $8\times8$ square and $r' = 2r$. (a)-(b) For the Ising Hamiltonian's CMI, we set $p_c = 0.178$, which is the analytically calculated value. It is marked by a black dashed line in panel (a). We then fit simulations with $\alpha = 2$ and $\nu = 1.1$ according to Eq.~\eqref{eq: finite_fit} and present them in panel (b). 
(c)-(d) For the Gauge Hamiltonian, by fitting simulation data we find that $p_c = 0.5$, $\alpha = 0.7$ and $\nu = 1.25$.}
\label{fig:figure 4}
\end{figure*}
\subsection{Numerical study of R\'enyi-2 CMI}
\label{sec: renyi-2}

The analysis of a decohered density matrix described above can be straightforwardly extended to any finite R\'enyi order.
In particular, transitions (if they exist) can be identified through nonanalyticities in the 
R\'enyi-$n$ CMI. To be concrete, here we focus on the case of $n=2$, for which a mixed-state transition
can be interpreted as a phase transition of the corresponding Choi state. In the present case, the decohered cluster state should exhibit the usual (single-flavor) Ising transition as expected from the mapping to the statistical model (cf. Eq.~\eqref{isingH}). To verify this analytical argument, we numerically compute the R\'enyi-2 CMI for the cluster state subjected to $\hat{X}$-noise described by Eq.~\eqref{eq: mixed state rho2}. We performed simulations for system sizes up to $92\times92$ adopting a tensor network method. We defer details of the numerical method to Appendix~\ref{sec: tensor network}. It should be noted that, unlike previous methods such as~\cite{Sang_2024_QMI}, which rely on extensive sampling of random noise configurations (up to $10^6$ samples), our method exactly computes the R\'enyi-2 CMI, thus drastically saving computational power.

We present the numerically computed $I_2(A:C|B)$ in Fig.~\ref{fig:figure 4}. To reduce the impact of finite size effects, it is advantageous to ensure that the size of region $A$ is not too small and that $r'\gg r$. Accordingly, we choose  $A$ as an $8\times8$ square and set $r'= 2r$ for $r\in\{6,8,10,12,14\}$. As previously argued in Ref.~\cite{Guo_2024}, the CMI of the entire cluster state can be understood as originating from two contributions: the Ising Hamiltonian and the gauge Hamiltonian. For clarity, we report and discuss these contributions separately in what follows. 

When $n = 2$, the contribution from the Ising Hamiltonian can be studied with the Kramers-Wannier duality~\cite{Kihara_1954} with a critical point at $p_{c}\simeq 0.178$ \cite{Guo_2024}. This result holds here as well, reflected in Fig.~\ref{fig:figure 4}(a): As we increase $r$, the position of the peak in the CMI approaches this value. Translated to the phonon parameters, the critical point is predicted to appear at $g/\omega\sim 0.4694$.

Next, to validate the finite-size scaling theory, we fit the R\'enyi-2 CMI to a power-law decaying function
\begin{equation}\label{eq: finite_fit}
     I_2(A:C|B)= r^{-\alpha}g(r^{\frac{1}{\nu}}(p-p_c)),
\end{equation}
where $p_c$ is set to $0.178$. In Fig.~\ref{fig:figure 4}(b), we numerically fit the lines and get $\alpha = 2$ and $\nu = 1.1$, where exponent of 2D Ising model is $\nu_{\text{Ising}}=1$.
As a comparison, we also perform additional simulations with varying sizes of the region $A$ and different ratios of $r'/r$, obtaining the following scaling factors:
\begin{enumerate}
    \item[$i)$] when region $A$ is set to $4\times4$ and $r'= 2r$, we find that $\alpha = 3.2,\nu = 1.1$;
    \item[$ii)$] when region  $A$ is set to $8\times8$ and $r'= r$, we find that $\alpha = 2,\nu = 1.15$.
\end{enumerate}

Furthermore, for the gauge Hamiltonian, we report results in Fig.~\ref{fig:figure 4}(c). At first glance, it appears that a phase-transition point can also be found, violating the theoretical prediction that would rule it out for the Gauge Hamiltonian Eq.~\eqref{gaugeH}. However, after fitting $p_c$, $\alpha$, and $\nu$, we find that $p_c=0.5$, which corresponds to the infinite-temperature limit in the statistical mechanics picture. This value does not suggest a meaningful phase transition. In summary, our numerical study validates the analytical results of Sec. \ref{sec: von neumann} obtained via a statistical mechanics mapping~\cite{Guo_2024}, both in predicting the existence of a single phase transition and in accounting for the finite size scaling effect.

\section{Discussion and Conclusions}
\label{sec: Discussion and conclusion}

In this work, we examined the phases of the 2D cluster Hamiltonian locally coupled to bosonic modes with arbitrary coupling strength. 
We first demonstrated that the standard pure-state approach---employing a disentangling unitary followed by a low energy manifold restriction---failed to capture the expected phase transition from an SPT phase to a trivial topological phase. Specifically, the pure state analysis showed that, irrespective of the spin-phonon coupling strength, the resulting spin Hamiltonian, dressed with bath parameters, retained the structure of the original Hamiltonian. Consequently, its ground-state wavefunction remained identical to that of the spin system without the coupled phonons.
Next, we adopted a mixed-state approach to interrogate the impact of phonons on the cluster state.
Explicitly, we traced out the phonons degrees of freedom {\it after} applying the entangling unitary (conjugate polaron transform), thus obtaining a phonon-decohered mixed state for the spin-Holstein type model. We found that the resulting mixed state corresponded to the pure 2D cluster state subjected to local bit-flip errors, with the error rate determined by the bath parameters. 

To diagnose the critical behavior of the obtained mixed state, we employed two diagnostic measures: von Neumann and Rényi-2 conditional mutual information. The von Neumann CMI is based on the two-way channel connectivity equivalence relation, and we analytically determined that a  critical behavior is expected at the RBIM critical point, given by $g/\omega\sim 0.352$ (Sec.~\ref{sec: von neumann}). The Rényi-2 CMI is based on the equivalence relation defined by the quasi-local unitary connection in the doubled Hilbert space. Using tensor network methods, we numerically identified the divergence of the Markov length, which occurred at the 2D classical Ising critical point. In this case, the critical spin-phonon coupling was given by $g/\omega\sim 0.4694$ (Sec.~\ref{sec: renyi-2}). Although the critical spin-phonon coupling strength occurs at the same order of magnitude as the phonon frequency $\omega$ for both diagnostics, their notions of phase transition are slightly different. Nevertheless, we expect mixed-state diagnostics based on equivalence relations to have the potential to advance our understanding of the notion of phases of matter in electron/spin-phonon systems. 

An important extension of our work is to establish a concrete connection between the divergent Markov length---both in the von Neumann and Rényi-2 senses---and the transition from strong-to-weak spontaneous symmetry breaking. Furthermore, while the pure-state approach failed to detect the phase transition in the spin-Holstein cluster model analyzed here, it would be valuable to explore strategies to overcome this limitation. 
For instance, it is a meaningful question to ask under what conditions a more sophisticated pure-state approach, such as an alternative choice to the variational wavefunction, might successfully identify the phase transition within the usual pure-state framework.

\section*{Acknowledgments}
We acknowledge useful discussions with Kazuki Yokomizo, Keisuke Fujii, Kanta Masuki, Jun Mochida, Masahiro Hoshino, and Yong-Baek Kim. The work of B.M. was supported by an JSPS-Mitacs Globalink Research Award and the Ontario Graduate Scholarship (OGS). Y.Z. acknowledges support from the Center for Quantum Materials and a CQIQC Fellowship at the University of Toronto, and thanks Juan Carrasquilla and ETH Zurich for their hospitality. Y.G. is financially supported by the Global Science Graduate Course (GSGC) program at the University of Tokyo. D.S.
acknowledges support from an NSERC Discovery Grant
and the Canada Research Chair program. Y.A. acknowledges support from the Japan Society for the Promotion of Science through Grant No.~JP19K23424 and from JST FOREST Program (Grant No.~JPMJFR222U, Japan). 

\appendix
\begin{widetext}
\section{Derivation for the effective system Hamiltonian}
\label{sec: Derivation for the effective system Hamiltonian}

In this appendix, we go through a detailed derivation for the effective system Hamiltonian provided in Eq.~\eqref{eq: effh}. 
We begin with the total Hamiltonian of the form: $\hat{H}=\hat{H}_S+\hat{H}_B+\hat{H}_I$, where
\begin{equation}
\begin{aligned}
    \hat{H}_S =& -\sum_v\hat{X}_v \hat{Z}_{e_1} \hat{Z}_{e_2} \hat{Z}_{e_3} \hat{Z}_{e_4}-\sum_e\hat{X}_e \hat{Z}_{v_1} \hat{Z}_{v_2},\\
    \hat{H}_I =& g\sum_v\hat{\tau}^x_v\left(\hat{b}^\dagger_v+\hat{b}_v\right)+g\sum_e\hat{\sigma}^x_e\left(\hat{b}^\dagger_e+\hat{b}_e\right),\\
    \hat{H}_B = &\omega \sum_v\hat{b}^\dagger_v\hat{b}_v+\omega\sum_e\hat{b}^\dagger_e\hat{b}_e.
    \end{aligned}
\end{equation}
From here, we apply the polaron unitary transform defined as
$\hat{U}_p=\prod_i\hat{U}_i$ where $\hat{U}_i=\exp(-\frac{i}{2}\hat{X}_i\hat{B}_i)$ and $\hat{B}_i=\frac{2ig}{\omega}(\hat{b}^\dagger_i-\hat{b}_i)$ to each terms in the Hamiltonian; the subscript $i$ covers both $v$ and $e$ points on the lattice. 
First, the spin system Hamiltonian transforms as  
\begin{equation}
\label{eq: system pol H 0}\hat{U}_p\hat{H}_S\hat{U}^\dagger_p=-\sum_v\hat{X}_v\prod^4_{j=1}\left(\hat{\mathcal{C}}_{e_j}\hat{Z}_{e_j}-\hat{\mathcal{S}}_{e_j}\hat{Y}_{e_j}\right)
  -\sum_e\hat{X}_e\prod^2_{j=1}\left(\hat{\mathcal{C}}_{v_j}\hat{Z}_{v_j}-\hat{\mathcal{S}}_{v_j}\hat{Y}_{v_j}\right).
\end{equation}
Here, we have used the short notation $\hat{\mathcal{C}}_{i}=\cos\left(\frac{2ig}{\omega}(\hat{b}^\dagger_i-\hat{b}_i)\right)$ and $\hat{\mathcal{S}}_i=\sin\left(\frac{2ig}{\omega}(\hat{b}^\dagger_i-\hat{b}_i)\right)$. 
Next, the interaction Hamiltonian $\hat{H}_I$ is transformed as 
\begin{equation}
    \hat{U}_p\hat{H}_I\hat{U}^\dagger_p=g\sum_v\hat{X}_v\left(\hat{b}^\dagger_v+\hat{b}_v\right)+g\sum_e\hat{X}_e\left(\hat{b}^\dagger_e+\hat{b}_e\right)-\frac{2Ng^2},{\omega}
\end{equation}
where $N$ is the total number of lattice sites (both vertices and edges). 
Lastly, the bath Hamiltonian $\hat{H}_B$ transforms as  
\begin{equation}
\hat{U}_p\hat{H}_B\hat{U}^\dagger_p= \omega \sum_v\hat{b}^\dagger_v\hat{b}_v+\omega\sum_v\hat{b}^\dagger_e\hat{b}_e-g\sum_v\hat{X}_v\left(\hat{b}^\dagger_v+\hat{b}_v\right)-g\sum_e\hat{X}_e\left(\hat{b}^\dagger_e+\hat{b}_e\right)+\frac{Ng^2}{\omega}.
\end{equation}
Combining all the transformed terms, we end up with 
\begin{equation}
\label{eq: polaron transformed H}\hat{U}_p\hat{H}\hat{U}^\dagger_p=-\sum_v\hat{X}_v\prod^4_{j=1}\left(\hat{\mathcal{C}}_{e_j}\hat{Z}_{e_j}-\hat{\mathcal{S}}_{e_j}\hat{Y}_{e_j}\right)
  -\sum_e\hat{X}_e\prod^2_{j=1}\left(\hat{\mathcal{C}}_{v_j}\hat{Z}_{v_j}-\hat{\mathcal{S}}_{v_j}\hat{Y}_{v_j}\right)+\omega\sum_v\hat{b}^\dagger_v\hat{b}_v+\omega\sum_e\hat{b}^\dagger_e\hat{b}_e-\frac{Ng^2}{\omega}.
\end{equation}
Before projecting the above Hamiltonian onto the vacuum modes of the harmonic oscillators, note that it retains the original $\mathds{Z}^{(0)}_2 \times \mathds{Z}^{(1)}_2$ symmetry of the 2D cluster Hamiltonian, defined by $\prod_{v \in A} \hat{X}_v$ and $\prod_{e \in \text{loop on B}} \hat{X}_e$. This retention occurs because the coupling between the lattice spins and their local harmonic displacement modes is mediated by $\hat{X}$ couplings, which commute with the original symmetries of the 2D cluster Hamiltonian. We emphasize this point because restricting to the low-energy manifold of the phonon degrees of freedom can sometimes artificially restore the original symmetry. For example, if the coupling to the harmonic modes were mediated via $\hat{Z}$ instead of $\hat{X}$—which breaks the original $\mathds{Z}^{(0)}_2 \times \mathds{Z}^{(1)}_2$ symmetry—the truncation to the ground-state phonon modes following the polaron transform would result in an artificial restoration of the symmetry.

From the polaron transform Hamiltonian described by Eq.~\eqref{eq: polaron transformed H}, we project to the ground state of the harmonic modes $\ket{0}=\otimes^N_{i=1}\ket{0_i}$. The only surviving terms inside the product in Eq.~\eqref{eq: system pol H 0} would be products of $\hat{\mathcal{C}}_i$ since $\hat{S}_i$ is odd in powers of $\hat{b}^\dagger_i-\hat{b}_i$. We are therefore left with
\begin{equation}
\label{eq: effective sys H 0}
    \hat{H}^\text{eff}_S=-\kappa_v\sum_v\hat{X}_v\hat{Z}_{e_1} \hat{Z}_{e_2} \hat{Z}_{e_3} \hat{Z}_{e_4}-\kappa_e\sum_e\hat{X}_e\hat{Z}_{v_1}\hat{Z}_{v_2}-\frac{Ng^2}{\omega}
\end{equation}
where
\begin{equation}
\label{eq: kappav kappae}
\begin{aligned}
\kappa_v = &\bra{0_{e_1}}\cos(\hat{B}_{e_1})\ket{0_{e_1}}\bra{0_{e_2}}\cos(\hat{B}_{e_2})\ket{0_{e_2}}\bra{0_{e_3}}\cos(\hat{B}_{e_3})\ket{0_{e_4}}\bra{0_{e_4}}\cos(\hat{B}_{e_4})\ket{0_{e_4}}=e^{-\frac{8g^2}{\omega^2}}\\
\kappa_e=&\bra{0_{v_1}}\cos(\hat{B}_{v_1})\ket{0_{v_1}}\bra{0_{v_2}}\cos(\hat{B}_{v_2})\ket{0_{v_2}}=e^{-\frac{4g^2}{\omega^2}}.
\end{aligned}
\end{equation}
The unique ground state of $\hat{H}^\text{eff}_S$, as given in Eq.~\eqref{eq: effective sys H 0}, remains the 2D cluster state with the SPT order. This is because the exponential suppression of the $\hat{A}_v$ and $\hat{B}_e$ terms does not alter the overall sign of these terms. As a result, the Hamiltonian retains its original form.

\section{Details on the mixed state derivation}
\label{sec: Details on the mixed state derivation}
In this appendix, we derive Eq.~\eqref{eq: mixed state rho2} in the main text. Recall that the ground state in the original frame can be written as 
\begin{equation}
    \ket{\Psi_\text{tot}}=\hat{U}\left(\ket{\tilde{\Psi}_\text{GS}}\otimes\ket{0}\right).
\end{equation}
where $\hat{U}=\hat{U}^\dagger_p$. We can therefore write 
\begin{equation}
\label{eq: x basis projection}
\begin{aligned}
    \hat{\rho}_2 =& \Tr_B[\ket{\Psi_\text{tot}}\bra{\Psi_\text{tot}}]\\
    =&\Tr_B\left[\hat{U}\left(\ket{\tilde{\Psi}_\text{GS}}\otimes\ket{0}\right)\left(\bra{\tilde{\Psi}_\text{GS}}\otimes\bra{0}\right)\hat{U}^\dagger\right]\\
    =&\prod_i\Big(\bra{+_i}\hat{\rho}_{2,i}\ket{+_i}\ket{+_i}\bra{+_i}+\bra{+_i}\hat{\rho}_{2,i}\ket{-_i}\ket{+_i}\bra{-_i}+\bra{-_i}\hat{\rho}_{2,i}\ket{+_i}\ket{-_i}\bra{+_i}+\bra{-_i}\hat{\rho}_{2,i}\ket{-_i}\ket{-_i}\bra{-_i}\Big),
    \end{aligned}
\end{equation}
where $\ket{+_i}$ and $\ket{-_i}$ are eigenstates of $\hat{X}_i$: $\hat{X}_i\ket{\pm_i}=\pm\ket{\pm_i}$. That is, we are projecting on to the eigenstates of $\hat{X}_i$. Each of the four terms under the product of Eq.~\eqref{eq: x basis projection} is given as
\begin{equation}
\label{eq: each projections}
    \begin{aligned}
\bra{+_i}\hat{\rho}_{2,i}\ket{+_i}\ket{+_i}\bra{+_i} = \Tr_B\left(\bra{+_i}e^{\frac{g}{\omega}(\hat{b}^\dagger_i-\hat{b}_i)}\ket{\tilde{\Psi}_\text{GS}}\otimes\ket{0_i}\bra{0_i}\otimes\bra{\tilde{\Psi}_\text{GS}}e^{-\frac{g}{\omega}(\hat{b}^\dagger_i-\hat{b}_i)}\ket{+_i}\right)\ket{+_i}\bra{+_i}\\
\bra{+_i}\hat{\rho}_{2,i}\ket{-_i}\ket{+_i}\bra{-_i} = \Tr_B\left(\bra{+_i}e^{\frac{g}{\omega}(\hat{b}^\dagger_i-\hat{b}_i)}\ket{\tilde{\Psi}_\text{GS}}\otimes\ket{0_i}\bra{0_i}\otimes\bra{\tilde{\Psi}_\text{GS}}e^{-\frac{g}{\omega}(\hat{b}^\dagger_i-\hat{b}_i)}\ket{-_i}\right)\ket{+_i}\bra{-_i}\\
\bra{-_i}\hat{\rho}_{2,i}\ket{+_i}\ket{-_i}\bra{+_i} = \Tr_B\left(\bra{-_i}e^{\frac{g}{\omega}(\hat{b}^\dagger_i-\hat{b}_i)}\ket{\tilde{\Psi}_\text{GS}}\otimes\ket{0_i}\bra{0_i}\otimes\bra{\tilde{\Psi}_\text{GS}}e^{-\frac{g}{\omega}(\hat{b}^\dagger_i-\hat{b}_i)}\ket{+_i}\right)\ket{-_i}\bra{+_i}\\
\bra{-_i}\hat{\rho}_{2,i}\ket{-_i}\ket{-_i}\bra{-_i} = \Tr_B\left(\bra{-_i}e^{\frac{g}{\omega}(\hat{b}^\dagger_i-\hat{b}_i)}\ket{\tilde{\Psi}_\text{GS}}\otimes\ket{0_i}\bra{0_i}\otimes\bra{\tilde{\Psi}_\text{GS}}e^{-\frac{g}{\omega}(\hat{b}^\dagger_i-\hat{b}_i)}\ket{-_i}\right)\ket{-_i}\bra{-_i}
    \end{aligned}
\end{equation}
where $\hat{X}_i$ in the exponential has already been acted on the eigenstate inside the trace. Hence, we are able to factor out the spin-wave function outside the trace. The first line of Eq.~\eqref{eq: each projections} can be therefore written as 
\begin{equation}
\bra{+_i}\hat{\rho}_{2,i}\ket{+_i}\ket{+_i}\bra{+_i} =\Tr_B\left(e^{\frac{g}{\omega}(\hat{b}^\dagger_i-\hat{b}_i)}\ket{0_i}\bra{0_i}e^{-\frac{g}{\omega}(\hat{b}^\dagger_i-\hat{b}_i)}\right)\ket{+_i}\bra{+_i}\ket{\tilde{\Psi}_\text{GS}}\bra{\tilde{\Psi}_\text{GS}}\ket{+_i}\bra{+_i} = \hat{P}_{+i}\hat{\rho}_1\hat{P}_{+i},
\end{equation}
where we have written the eigenstate projectors as $\hat{P}_{\pm i}=\ket{\pm_i}\bra{\pm_i}=\frac{1\pm\hat{X}_i}{2}$. Similarly, for the last line of Eq.~\eqref{eq: each projections}, we get  
\begin{equation}
    \bra{-_i}\hat{\rho}_{2,i}\ket{-_i}\ket{-_i}\bra{-_i} =\Tr_B\left(e^{-\frac{g}{\omega}(\hat{b}^\dagger_i-\hat{b}_i)}\ket{0_i}\bra{0_i}e^{+\frac{g}{\omega}(\hat{b}^\dagger_i-\hat{b}_i)}\right)\ket{-_i}\bra{-_i}\ket{\tilde{\Psi}_\text{GS}}\bra{\tilde{\Psi}_\text{GS}}\ket{-_i}\bra{-_i} = \hat{P}_{-i}\hat{\rho}_1\hat{P}_{-i}.
\end{equation}
It is the cross terms that get modified by an exponential suppression. For instance, note for the second line of 
Eq.~\eqref{eq: each projections},
\begin{equation}
\begin{aligned}
    \bra{+_i}\hat{\rho}_{2,i}\ket{-_i}\ket{+_i}\bra{-_i}=&\Tr_B\left(\bra{+_i}e^{\frac{g}{\omega}(\hat{b}^\dagger_i-\hat{b}_i)}\ket{\tilde{\Psi}_\text{GS}}\otimes\ket{0_i}\bra{0_i}\otimes\bra{\tilde{\Psi}_\text{GS}}e^{+\frac{g}{\omega}(\hat{b}^\dagger_i-\hat{b}_i)}\ket{-_i}\right)\ket{+_i}\bra{-_i}\\
    =& \Tr_B\left(e^{\frac{g}{\omega}(\hat{b}^\dagger_i-\hat{b}_i)}\ket{0_i}\bra{0_i}e^{\frac{g}{\omega}(\hat{b}^\dagger_i-\hat{b}_i)}\right)\ket{+_i}\bra{+_i}\ket{\tilde{\Psi}_\text{GS}}\bra{\tilde{\Psi}_\text{GS}}\ket{-_i}\bra{-_i}\\
    =&\sum_m\bra{m_i}\ket{0_i}\bra{0_i}e^{\frac{2g}{\omega}(\hat{b}^\dagger_i-\hat{b}_i)}\ket{m_i}\hat{P}_{+i}\hat{\rho}_1\hat{P}_{-i}\\
    =&\bra{0_i}e^{\frac{2g}{\omega}(\hat{b}^\dagger_i-\hat{b}_i)}\ket{0_i}\hat{P}_{+i}\hat{\rho}_1\hat{P}_{-i}=e^{-\frac{2g^2}{\omega^2}}\hat{P}_{+i}\hat{\rho}_1\hat{P}_{-i}.
    \end{aligned}
\end{equation}
One can easily check that the same factor of $e^{-\frac{2g^2}{\omega^2}}$ arises on the third line of Eq.~\eqref{eq: each projections}. We therefore obtain
\begin{equation}
    \hat{\rho}_2=\mathcal{E}_{2\leftarrow 1}[\hat{\rho}_1]=\prod_i\left(\hat{P}_{+i}\hat{\rho}_1\hat{P}_{+i}+\hat{P}_{-i}\hat{\rho}_1\hat{P}_{-i}+e^{-{2\left(\frac{g}{\omega}\right)^2}}\left(\hat{P}_{+i}\hat{\rho}_1\hat{P}_{-i}+\hat{P}_{-i}\hat{\rho}_1\hat{P}_{+i}\right)\right).
\end{equation}
Noting 
\begin{equation}
    \begin{cases}
        \hat{P}_{+i}\hat{\rho}_1\hat{P}_{+i}+\hat{P}_{-i}\hat{\rho}_1\hat{P}_{-i} = \frac{1}{2}\left[\hat{\rho}_1+\hat{X}_i\hat{\rho}_1\hat{X}_i\right]\\
       \hat{P}_{+i}\hat{\rho}_1\hat{P}_{-i}+\hat{P}_{-i}\hat{\rho}_1\hat{P}_{+i} = \frac{1}{2}\left[\hat{\rho}_1-\hat{X}_i\hat{\rho}_1\hat{X}_i\right] 
    \end{cases},
\end{equation}
we are finally able to write 
\begin{equation}
    \hat{\rho}_2=\mathcal{E}_{2\leftarrow 1}[\hat{\rho}_1]=\prod_i\left(\frac{1}{2}\left[1+e^{-2(\frac{g}{\omega})^2}\right]\hat{\rho}_1+\frac{1}{2}\left[1-e^{-2(\frac{g}{\omega})^2}\right]\hat{X}_i\hat{\rho}_1\hat{X}_i\right).
\end{equation}
This describes a quantum channel that takes a pure state of the spin system, $\hat{\rho}_1=\ket{\tilde{\Psi}_\text{GS}}\bra{\tilde{\Psi}_\text{GS}}$, and transform it to a mixed state, $\hat{\rho}_2$, with local bit-flip error of rate $\frac{1}{2}(1-e^{-2(\frac{g}{\omega})^2})$.

\section{Details on the variational process}
\label{sec: variational process}

Instead of performing a ``full" polaron transform, we can define the polaron unitary as $\hat{U}_\lambda = \prod_i \hat{U}_i(\lambda)$, where $\hat{U}_i(\lambda) = e^{\frac{\lambda}{\omega} \hat{X}_i (\hat{b}^\dagger_i - \hat{b}_i)}$. $\lambda$ is a variational parameter, replacing $g$, the spin-phonon coupling energy parameter.
Additionally, we may choose the bosonic variational wave function to be a coherent state, $\hat{b}_i \ket{\alpha_i} = \alpha \ket{\alpha_i}$, instead of the vacuum state of the bosonic modes. We then determine the optimal values for $\lambda$ and $\alpha = x + ip$ by minimizing the corresponding energy functional,
\begin{equation}   E(\lambda,x,p)=\bra{\Psi_\text{tot}}\hat{H}\ket{\Psi_\text{tot}}= (\bra{\alpha}\otimes\bra{\tilde{\Psi}_\text{GS}})\hat{U}_\lambda \hat{H}\hat{U}^\dagger_\lambda(\ket{\tilde{\Psi}_\text{GS}}\otimes\ket{\alpha}),
\end{equation}
where the minimization is performed with respect to $\lambda$, $x$, and $p$. We assume a coherent state as the state of each mode, and construct $\ket{\alpha}=\otimes_i\ket{\alpha_i}$.
We now transform the Hamiltonian with this variational transformation,
writing $\hat{U}_\lambda \hat{H} \hat{U}^\dagger_\lambda$, as 
\begin{equation}
\label{eq: polaron transform lam}
\begin{aligned}\hat{U}_\lambda\hat{H}\hat{U}^\dagger_\lambda=-&\sum_v\hat{X}_v\left(\hat{\mathcal{C}}_{e_1}\hat{Z}_{e_1}-\hat{\mathcal{S}}_{e_1}\hat{Y}_{e_1}\right)\left(\hat{\mathcal{C}}_{e_2}\hat{Z}_{e_2}-\hat{\mathcal{S}}_{e_2}\hat{Y}_{e_2}\right)\left(\hat{\mathcal{C}}_{e_3}\hat{Z}_{e_3}-\hat{\mathcal{S}}_{e_3}\hat{Y}_{e_3}\right)\left(\hat{\mathcal{C}}_{e_4}\hat{Z}_{e_4}-\hat{\mathcal{S}}_{e_4}\hat{Y}_{e_4}\right)
  \\
  -&\sum_e\hat{X}_e\left(\hat{\mathcal{C}}_{v_2}\hat{Z}_{v_2}-\hat{\mathcal{S}}_{v_2}\hat{Y}_{v_2}\right)\left(\hat{\mathcal{C}}_{v_2}\hat{Z}_{v_2}-\hat{\mathcal{S}}_{v_2}\hat{Y}_{v_2}\right)\\
  +&(g-\lambda)\sum_v\hat{X}_v\left(\hat{b}^\dagger_v+\hat{b}_v\right)+(g-\lambda)\sum_e\hat{X}_e\left(\hat{b}^\dagger_e+\hat{b}_e\right)\\
+&\omega\sum_v\hat{b}^\dagger_v\hat{b}_v+\omega\sum_e\hat{b}^\dagger_e\hat{b}_e-\frac{\lambda(\lambda-2g)}{\omega}N.
  \end{aligned}
\end{equation}
Here, $\hat{\mathcal{C}}_i = \cos\left(\frac{2i\lambda}{\omega} (\hat{b}^\dagger_i - \hat{b}_i)\right)$ and $\hat{\mathcal{S}}_i = \sin\left(\frac{2i\lambda}{\omega} (\hat{b}^\dagger_i - \hat{b}_i)\right)$. Here, $\lambda$ rather than $g$ defines the dressing functions, contrasting from the full polaron treatment. 
Notably, even under the coherent state projection, none of the terms involving $\hat{\mathcal{S}}i$ in the first two lines of Eq.~\eqref{eq: polaron transform lam} contributes. This is due to the symmetry of our variational wave function $\ket{\tilde{\Psi}_\text{GS}}$, which is the unique ground state of the original 2D cluster Hamiltonian. Specifically, acting with $\hat{A}_v$ or $\hat{B}_e$ for any vertex $v$ or edge $e$ on $\ket{\tilde{\Psi}_\text{GS}}$ results in at most a phase factor, which is known to be $1$:
\begin{equation}
    \begin{cases}
        \hat{A}_v\ket{\tilde{\Psi}_\text{GS}}=e^{i\theta_A}\ket{\tilde{\Psi}_\text{GS}}&\text{for all $v$},\\
    \hat{B}_e\ket{\tilde{\Psi}_\text{GS}}=e^{i\theta_B}\ket{\tilde{\Psi}_\text{GS}}&\text{for all $e$}.\\
    \end{cases}
\end{equation}
Hence, if an operator that we wish to compute an average of, anti-commutes with any $\hat{A}_v$ or $\hat{B}_e$, we get zero:
\begin{equation}
\begin{cases}\bra{\tilde{\Psi}_\text{GS}}\hat{\mathcal{O}}\ket{\tilde{\Psi}_\text{GS}}=\bra{\tilde{\Psi}_\text{GS}}\hat{A}_v\hat{\mathcal{O}}\hat{A}_v\ket{\tilde{\Psi}_\text{GS}}=-\bra{\tilde{\Psi}_\text{GS}}\hat{\mathcal{O}}\ket{\tilde{\Psi}_\text{GS}}=0& \text{for any $v$ where $\{\hat{\mathcal{O}},\hat{A}_v\}=0$},\\
\bra{\tilde{\Psi}_\text{GS}}\hat{\mathcal{O}}\ket{\tilde{\Psi}_\text{GS}}=\bra{\tilde{\Psi}_\text{GS}}\hat{B}_e\hat{\mathcal{O}}\hat{B}_e\ket{\tilde{\Psi}_\text{GS}}=-\bra{\tilde{\Psi}_\text{GS}}\hat{\mathcal{O}}\ket{\tilde{\Psi}_\text{GS}}=0& \text{for any $e$ where $\{\hat{\mathcal{O}},\hat{B}_e\}=0$}.
\end{cases}
\end{equation}
For example, we can immediately see that 
\begin{equation}
\begin{aligned}
    \bra{\tilde{\Psi}_\text{GS}}
    \hat{X}_v\hat{Y}_{e_1}\hat{Y}_{e_2}\hat{Y}_{e_3}\hat{Y}_{e_4}
    \ket{\tilde{\Psi}_\text{GS}}=&\bra{\tilde{\Psi}_\text{GS}}
    (\hat{X}_{v'}\hat{Z}_{e_1}\hat{Z}_{e_{2'}}\hat{Z}_{e_{3'}}\hat{Z}_{e_{4'}})\hat{X}_v\hat{Y}_{e_1}\hat{Y}_{e_2}\hat{Y}_{e_3}\hat{Y}_{e_4}(\hat{X}_{v'}\hat{Z}_{e_1}\hat{Z}_{e_{2'}}\hat{Z}_{e_{3'}}\hat{Z}_{e_{4'}})\ket{\tilde{\Psi}_\text{GS}}\\
    =&- \bra{\tilde{\Psi}_\text{GS}}
    \hat{X}_v\hat{Y}_{e_1}\hat{Y}_{e_2}\hat{Y}_{e_3}\hat{Y}_{e_4}
    \ket{\tilde{\Psi}_\text{GS}}=0.
    \end{aligned}
\end{equation}
Under the same reasoning, other terms involving $\hat{S}_i$ can be shown to vanish once we take the expectation value with respect to $\ket{\tilde{\Psi}_\text{GS}}$, as well as $\bra{\tilde{\Psi}_\text{GS}}\hat{X}_i\ket{\tilde{\Psi}_\text{GS}}=0$. After evaluating the expectation values on the system wave function, we are left with 
\begin{equation}
\begin{aligned}E(\lambda,x,p)=&\bra{\Psi_\text{tot}}\hat{H}\ket{\Psi_\text{tot}}\\
=&-\sum_v\left(\bra{\alpha_{e}}\hat{\mathcal{C}}_e\ket{\alpha_e}\right)^4-\sum_e\left(\bra{\alpha_{v}}\hat{\mathcal{C}}_v\ket{\alpha_v}\right)^2+\omega N|\alpha|^2+\left[\frac{\lambda(\lambda-2g)}{\omega}\right]N.
\end{aligned}
\end{equation}
Let us now evaluate $\bra{\alpha_i}\hat{\mathcal{C}}_i\ket{\alpha_i}$. First we write
\begin{equation}
    \bra{\alpha_i}\hat{\mathcal{C}}_i\ket{\alpha_i}=\bra{\alpha_i}\frac{\exp\left[-2\frac{\lambda}{\omega}\left(\hat{b}^\dagger_i-\hat{b}_i\right)\right]+\exp\left[+2\frac{\lambda}{\omega}\left(\hat{b}^\dagger_i-\hat{b}_i\right)\right]}{2}\ket{\alpha_i}%
\end{equation}
Note that by Baker-Campbell-Hausdorff (BCH) formula,
\begin{equation}
\begin{aligned}
    e^{-2\frac{\lambda}{\omega}\left(\hat{b}^\dagger-\hat{b}\right)} =& e^{-2\frac{\lambda}{\omega}\hat{b}^\dagger}e^{+2\frac{\lambda}{\omega}\hat{b}}e^{-2\frac{\lambda^2}{\omega^2}}\\
    e^{+2\frac{\lambda}{\omega}\left(\hat{b}^\dagger-\hat{b}\right)} =& e^{+2\frac{\lambda}{\omega}\hat{b}^\dagger}e^{-2\frac{\lambda}{\omega}\hat{b}}e^{-2\frac{\lambda^2}{\omega^2}}.
    \end{aligned}
\end{equation}
Therefore,
\begin{equation}
\begin{aligned}
\bra{\alpha_i}\hat{\mathcal{C}}_i\ket{\alpha_i}=&\frac{e^{-\frac{2\lambda^2}{\omega^2}}}{2}\left(\bra{\alpha_i}e^{-\frac{2\lambda}{\omega}\hat{b}^\dagger_i}e^{+\frac{2\lambda}{\omega}\hat{b}_i}\ket{\alpha_i}+\bra{\alpha_i}e^{+\frac{2\lambda}{\omega}\hat{b}^\dagger_i}e^{-\frac{2\lambda}{\omega}\hat{b}_i}\ket{\alpha_i}\right)\\
=&e^{-\frac{2\lambda^2}{\omega^2}}\frac{\exp({-\frac{2\lambda \alpha^*}{\omega}})\exp({+\frac{2\lambda\alpha}{\omega}})+\exp(+\frac{2\lambda \alpha^*}{\omega})\exp(-\frac{2\lambda \alpha}{\omega})}{2}\\
=&e^{-\frac{2\lambda^2}{\omega^2}}\frac{\exp(-\frac{2\lambda}{\omega}(\alpha^*-\alpha))+\exp(+\frac{2\lambda}{\omega}(\alpha^*-\alpha))}{2}\\
=&e^{-\frac{2\lambda^2}{\omega^2}}\cos\Big(\frac{4\lambda p}{\omega}\Big)
\end{aligned}
\end{equation}
Collecting all terms we get
\begin{equation}
    E(\lambda,x,p)=N_v\left(-e^{-\frac{8\lambda^2}{\omega^2}}\cos^4\Big(\frac{4\lambda}{\omega}p\Big)-2e^{-\frac{4\lambda^2}{\omega^2}}\cos^2\Big(\frac{4\lambda}{\omega}p\Big)+3w\left(x^2+p^2\right)+\frac{3\lambda(\lambda-2g)}{\omega}\right)
    \label{eq:EV}
\end{equation}
where $N_v$ is the number of vertex in the system. We immediately note that the optima value for $x=0$. Next, the optimal value for $p$ is dictated by 
\begin{equation}
     \frac{\partial E}{\partial p}=+4e^{-\frac{8\lambda^2}{\omega^2}}\cos^3\Big(\frac{4\lambda}{\omega}p\Big)\sin\Big(\frac{4\lambda}{\omega}p\Big)\frac{4\lambda}{\omega}+4e^{-\frac{4\lambda^2}{\omega^2}}\cos\Big(\frac{4\lambda}{\omega}\Big)\sin\Big(\frac{4\lambda}{\omega}p\Big)\frac{4\lambda}{\omega}+6\omega p=0
\end{equation}
Simplifying the equation, we have 
\begin{equation}
    p = -\frac{4\lambda}{3\omega^2}e^{-\frac{4\lambda^2}{\omega^2}}\sin\Big(\frac{8\lambda}{\omega}p\Big)\left[e^{-\frac{4\lambda^2}{\omega^2}}\cos^2\Big(\frac{4\lambda}{\omega}p\Big)+1\right].
\end{equation}
For wide range of $\lambda$, we find that $p=0$.
We conclude from the variational treatment that the general coherent state reduces to the ground state as in the standard ground state truncation scheme. As for the value of $\lambda$, one can optimize it as well and find a bath-dressed parameter for $\lambda$. However, this will not affect the overall structure of the effective spin-Hamiltonian as described in Eq.~\eqref{eq: effective sys H 0} therefore its ground state. That is, it will simply replace the argument of the exponential dressing parameters from $g$ to $\lambda$ in Eq.~\eqref{eq: kappav kappae},

We now address the following question: Suppose there were a non-trivial $\alpha$ that lowered the ground-state energy. What would be the corresponding error channel? We derive the error channel when the pure-state fed into the channel is a product state of $\ket{\tilde{\Psi}_\text{GS}}$ and $\ket{\alpha}$. That is, we assume that the ground state wave function of the polaron rotated Hamiltonian is
\begin{equation}
    \ket{\tilde{\Psi}_\text{tot}} = \ket{\tilde{\Psi}_\text{GS}}\otimes \ket{\alpha}.
\end{equation}
Here $\ket{\alpha}$ is the bosonic coherent state
\begin{equation}
   \ket{\alpha}=\ket{\alpha_{v_1}}\otimes\dots\otimes\ket{\alpha_{v_{N_v}}}\otimes\ket{\alpha_{e_1}}\otimes\dots\otimes\ket{\alpha_{v_{N_e}}},\quad\text{where} \quad \hat{b}_i\ket{\alpha_i}=\alpha_i\ket{\alpha_i}. 
\end{equation}
Following the same derivation as in Appendix~\ref{sec: Details on the mixed state derivation}, we now go through the mixed state generation by tracing out the bath's degrees of freedom. 
Analogous to Eq.~\eqref{eq: x basis projection}, we begin with 
\begin{equation}
\label{eq: x projection alpha}
\begin{aligned}
    \hat{\rho}_2 =& \Tr_B[\ket{\Psi_\text{tot}}\bra{\Psi_\text{tot}}]\\
    =&\Tr_B\left[\hat{U}\left(\ket{\tilde{\Psi}_\text{GS}}\otimes\ket{\mathbf{\alpha}}\right)\left(\bra{\tilde{\Psi}_\text{GS}}\otimes\bra{\mathbf{\alpha}}\right)\hat{U}^\dagger\right]\\
    =&\prod_i\Big(\bra{+_i}\hat{\rho}_{2,i}\ket{+_i}\ket{+_i}\bra{+_i}+\bra{+_i}\hat{\rho}_{2,i}\ket{-_i}\ket{+_i}\bra{-_i}+\bra{-_i}\hat{\rho}_{2,i}\ket{+_i}\ket{-_i}\bra{+_i}+\bra{-_i}\hat{\rho}_{2,i}\ket{-_i}\ket{-_i}\bra{-_i}\Big).
    \end{aligned}
\end{equation}
Computing each term under the product sign of Eq.~\eqref{eq: x projection alpha}, starting with the first term, we obtain
\begin{equation}
\bra{+_i}\hat{\rho}_{2,i}\ket{+_i}\ket{+_i}\bra{+_i} = \Tr_B\left(\bra{+_i}e^{\frac{\lambda}{\omega}(\hat{b}^\dagger_i-\hat{b}_i)}\ket{\tilde{\Psi}_\text{GS}}\otimes\ket{\alpha_i}\bra{\alpha_i}\otimes\bra{\tilde{\Psi}_\text{GS}}e^{-\frac{\lambda}{\omega}(\hat{b}^\dagger_i-\hat{b}_i)}\ket{+_i}\right)\ket{+_i}\bra{+_i}.
\end{equation}
As in Appendix.~\ref{sec: Details on the mixed state derivation}, we again have $\hat{X}_i$ in the exponential already acted on the eigenstate. This leads to 
\begin{equation}
    \Tr_B\left(e^{\frac{\lambda}{\omega}(\hat{b}^\dagger_i-\hat{b}_i)}\ket{\alpha_i}\bra{\alpha_i}e^{-\frac{\lambda}{\omega}(\hat{b}^\dagger_i-\hat{b}_i)}\right)\ket{+_i}\bra{+_i}\ket{\tilde{\Psi}_\text{GS}}\bra{\tilde{\Psi}_\text{GS}}\ket{+_i}\bra{+_i} = \hat{P}_{+i}\hat{\rho}_1\hat{P}_{+i}
\end{equation}
It is again only the cross terms that will get exponential factors. For instance, consider
\begin{equation}
   \bra{+_i}\hat{\rho}_{2,i}\ket{-_i}\ket{+_i}\bra{-_i}=\Tr_B\left(\bra{+_i}e^{\frac{\lambda}{\omega}(\hat{b}^\dagger_i-\hat{b}_i)}\ket{\tilde{\Psi}_\text{GS}}\otimes\ket{\alpha_i}\bra{\alpha_i}\otimes\bra{\tilde{\Psi}_\text{GS}}e^{+\frac{\lambda}{\omega}(\hat{b}^\dagger_i-\hat{b}_i)}\ket{-_i}\right)\ket{+_i}\bra{-_i},
\end{equation}
where again, $\hat{X}_i\ket{\pm_i}=\pm\ket{\pm_i}$ has already been applied to the $\hat{X}$ eigenstates. This leads to  
\begin{equation}
\label{eq: cross term alphha}
\begin{aligned}
    \bra{+_i}\hat{\rho}_{2,i}\ket{-_i}\ket{+_i}\bra{-_i}=&\Tr_B\left(e^{\frac{\lambda}{\omega}(\hat{b}^\dagger_i-\hat{b}_i)}\ket{\alpha_i}\bra{\alpha_i}e^{\frac{\lambda}{\omega}(\hat{b}^\dagger_i-\hat{b}_i)}\right)\ket{+_i}\bra{-_i}\\
    =&\sum_m\bra{m_i}\ket{\alpha_i}\bra{\alpha_i}e^{\frac{2\lambda}{\omega}(\hat{b}^\dagger_i-\hat{b}_i)}\ket{m_i}\ket{+_i}\bra{-_i}\\
    =& \bra{\alpha_i}e^{\frac{2\lambda}{\omega}(\hat{b}^\dagger_i-\hat{b}_i)}\ket{\alpha_i}\ket{+_i}\bra{-_i}.
    \end{aligned}
\end{equation}
Employing the BCH formula, we are able to write 
\begin{equation}
    e^{\frac{2\lambda}{\omega}\hat{b}^\dagger_i}e^{-\frac{2\lambda}{\omega}\hat{b}_i} = e^{\frac{2\lambda}{\omega}\left(\hat{b}^\dagger_i-\hat{b}_i\right)+\frac{1}{2}\left[\frac{2\lambda}{\omega}\hat{b}^\dagger_i,-\frac{2\lambda}{\omega}\hat{b}_i\right]+\dots}=e^{\frac{2\lambda}{\omega}\left(\hat{b}^\dagger_i-\hat{b}_i\right)}e^{+\frac{2\lambda^2}{\omega^2}}.
\end{equation}
And so, the matrix element in the last line of Eq.~\eqref{eq: cross term alphha} can be written as 
\begin{equation}
    \bra{\alpha_i}e^{\frac{2\lambda}{\omega}\left(\hat{b}^\dagger_i-\hat{b}_i\right)}\ket{\alpha_i}=e^{-\frac{2\lambda^2}{\omega^2}}\bra{\alpha_i} e^{\frac{2\lambda}{\omega}\hat{b}^\dagger_i}e^{-\frac{2\lambda}{\omega}\hat{b}_i}\ket{\alpha_i}=e^{-\frac{2\lambda^2}{\omega^2}}e^{\frac{2\lambda}{\omega}\alpha^*}e^{-\frac{2\lambda}{\omega}\alpha}=e^{-\frac{2\lambda^2}{\omega^2}}e^{\frac{2\lambda}{\omega}(\alpha^*-\alpha)}\bra{\alpha_i}\ket{\alpha_i}.
\end{equation}
The remaining cross term in Eq.~\eqref{eq: x projection alpha} can be written as
\begin{equation}
\begin{aligned}
   \bra{-_i}\hat{\rho}_{2,i}\ket{+_i}\ket{-_i}\bra{+_i} =&\Tr_B\left(\bra{-_i}e^{\frac{\lambda}{\omega}\hat{X}_i(\hat{b}^\dagger_i-\hat{b}_i)}\ket{\tilde{\Psi}_\text{GS}}\otimes\ket{\alpha_i}\bra{\alpha_i}\otimes\bra{\tilde{\Psi}_\text{GS}}e^{-\frac{\lambda}{\omega}\hat{X}_i(\hat{b}^\dagger_i-\hat{b}_i)}\ket{+_i}\right)\ket{-_i}\bra{+_i}\\
   =&\Tr_B\left(e^{-\frac{\lambda}{\omega}(\hat{b}^\dagger_i-\hat{b}_i)}\ket{\alpha_i}\bra{\alpha_i}e^{-\frac{\lambda}{\omega}(\hat{b}^\dagger_i-\hat{b}_i)}\right)\bra{-_i}\ket{\tilde{\Psi}_\text{GS}}\bra{\tilde{\Psi}_\text{GS}}\ket{+_i}\ket{-_i}\bra{+_i}\\
   =&\sum_m\bra{m_i}\ket{\alpha_i}\bra{\alpha_i}e^{-\frac{2\lambda}{\omega}(\hat{b}^\dagger_i-\hat{b}_i)}\ket{m_i} \ket{-_i}\bra{+_i}\\
   =&\bra{\alpha_i}^{-\frac{2\lambda}{\omega}(\hat{b}^\dagger_i-\hat{b}_i)}\ket{\alpha_i}\ket{-_i}\bra{+_i}.
   \end{aligned}
\end{equation}
Employing the BCH formula, we are able to write
\begin{equation}
    e^{-\frac{2\lambda}{\omega}\hat{b}^\dagger_i}e^{+\frac{2\lambda}{\omega}\hat{b}_i} = e^{-\frac{2\lambda}{\omega}(\hat{b}^\dagger_i-\hat{b}_i)+\frac{1}{2}\left[-\frac{2\lambda}{\omega}\hat{b}^\dagger_i,\frac{2\lambda}{\omega}\hat{b}_i\right]+\dots}=e^{-\frac{2\lambda}{\omega}(\hat{b}^\dagger_i-\hat{b}_i)}e^{+\frac{2\lambda^2}{\omega^2}}.
\end{equation}
Overall, we get
\begin{equation}
\begin{aligned}
    \bra{-_i}\hat{\rho}_{2,i}\ket{+_i}\ket{-_i}\bra{+_i}=&\bra{\alpha_i}e^{-\frac{2\lambda}{\omega}(\hat{b}^\dagger_i-\hat{b}_i)}\ket{\alpha_i}\ket{-_i}\bra{+_i}\\
    =&e^{-\frac{2\lambda^2}{\omega^2}}\bra{\alpha_i}e^{-\frac{2\lambda}{\omega}\hat{b}^\dagger_i}e^{+\frac{2\lambda}{\omega}\hat{b}_i} \ket{\alpha_i}\ket{-_i}\bra{+_i}\\
    =& e^{-\frac{2\lambda^2}{\omega^2}}e^{-\frac{2\lambda}{\omega}(\alpha^*-\alpha)}\bra{\alpha_i}\ket{\alpha_i}\ket{-_i}\bra{+_i}.
    \end{aligned}
\end{equation}
Eq.~\eqref{eq: x projection alpha} therefore reduces to
\begin{equation}
\label{eq: q channel alphha}
    \mathcal{E}_{2\leftarrow 1}(\hat{\rho}_1) = \prod_i\left(\hat{P}_{+i}\hat{\rho}_1\hat{P}_{+i}+\hat{P}_{-i}\hat{\rho}_1\hat{P}_{-i}+e^{-{2\left(\frac{\lambda}{\omega}\right)^2}}\left(e^{\frac{2\lambda}{\omega}(\alpha^*-\alpha)}\hat{P}_{+i}\hat{\rho}_1\hat{P}_{-i}+e^{-\frac{2\lambda}{\omega}(\alpha^*-\alpha)}\hat{P}_{-i}\hat{\rho}_1\hat{P}_{+i}\right)\right).
\end{equation}
Writing $\alpha=x+ip$, we have 
\begin{equation}
\begin{aligned}
    e^{\frac{2\lambda}{\omega}(\alpha^*-\alpha)} =& e^{\frac{2\lambda}{\omega}(-2ip)} =e^{-i\frac{4\lambda}{\omega}p}=\gamma\\
    e^{-\frac{2\lambda}{\omega}(\alpha^*-\alpha)} =& e^{-\frac{2\lambda}{\omega}(-2ip)} =e^{+i\frac{4\lambda}{\omega}p}=\gamma^*.
    \end{aligned}
\end{equation}
Recall that the terms free of bath parameters in Eq.~\eqref{eq: q channel alphha} can be written as 
$    \hat{P}_{+i}\hat{\rho}_1\hat{P}_{+i}+\hat{P}_{-i}\hat{\rho}_1\hat{P}_{-i} = \frac{1}{2}\left[\hat{\rho}_1+\hat{X}_i\hat{\rho}_1\hat{X}_i\right]
$. Terms that depend on the bath parameters are given as
\begin{equation}
\begin{aligned}
\gamma\hat{P}_{+i}\hat{\rho}_1\hat{P}_{-i}+\gamma^*\hat{P}_{-i}\hat{\rho}_1\hat{P}_{+i} =& \gamma \left(\frac{1+\hat{X_i}}{2}\right)\hat{\rho}_1 \left(\frac{1-\hat{X_i}}{2}\right)+\gamma^*\left(\frac{1-\hat{X_i}}{2}\right)\hat{\rho}_1 \left(\frac{1+\hat{X_i}}{2}\right)\\
=&\frac{\gamma}{4}\left[\hat{\rho}_1+\hat{X}_i\hat{\rho}_1-\hat{\rho}_1\hat{X}_i-\hat{X}_i\hat{\rho}\hat{X}_i\right]+\frac{\gamma^*}{4}\left[\hat{\rho}_1-\hat{X}_i\hat{\rho}_1+\hat{\rho}_1\hat{X}_i-\hat{X}_i\hat{\rho}\hat{X}_i\right]\\
=&\frac{\gamma+\gamma^*}{4}\hat{\rho}_1+\frac{\gamma-\gamma^*}{4}\hat{X}_i\hat{\rho}_1+\frac{\gamma^*-\gamma}{4}\hat{\rho}_1\hat{X}_i-\frac{\gamma+\gamma^*}{4}\hat{X}_i\hat{\rho}_1\hat{X}_i\\
=&\frac{1}{2}\cos\Big(\frac{4\lambda p}{\omega}\Big)\left[\hat{\rho}_1-\hat{X}_i\hat{\rho}_1\hat{X}_i\right]-\frac{i}{2}\sin\Big(\frac{4\lambda p}{\omega}\Big)\left[\hat{X}_i,\hat{\rho}_1\right].
\end{aligned}
\end{equation}
Combining the two contributions, we have
\begin{equation}
    \mathcal{E}_{2\leftarrow 1}(\hat{\rho}_1)=\prod_i\Big(\frac{1}{2}\left[\hat{\rho}_1+\hat{X}_i\hat{\rho}_1\hat{X}_i\right]+\frac{ e^{-\frac{2\lambda^2}{\omega^2}}}{2}\cos\Big(\frac{4\lambda p}{\omega}\Big)\left[\hat{\rho}_1-\hat{X}_i\hat{\rho}_1\hat{X}_i\right]-\frac{i e^{-\frac{2\lambda^2}{\omega^2}}}{2}\sin\Big(\frac{4\lambda p}{\omega}\Big)\left[\hat{X}_i,\hat{\rho}_1\right]\Big).
\end{equation}
Reorganizing, we get
\begin{equation}
    \mathcal{E}_{2\leftarrow 1}(\hat{\rho}_1)=\prod_i\Big(\frac{1}{2}\left[1+e^{-\frac{2\lambda^2}{\omega^2}}\cos\Big(\frac{4\lambda p}{\omega}\Big)\right]\hat{\rho}_1+\frac{1}{2}\left[1-e^{-\frac{2\lambda^2}{\omega^2}}\cos\Big(\frac{4\lambda p}{\omega}\Big)\right]\hat{X}_i\hat{\rho}_1\hat{X}_i-\frac{i e^{-\frac{2\lambda^2}{\omega^2}}}{2}\sin\Big(\frac{4\lambda p}{\omega}\Big)\left[\hat{X}_i,\hat{\rho}_1\right]\Big).
    \label{eq:Channel}
\end{equation}
It is useful to recast the above channel into a diagonal representation (which is always possible for a channel that transforms density matrix to another density matrix), so that we can easily classify the specific types of errors occurring. That is, we represent the channel in the following form
\begin{equation}
    \mathcal{E}_{2\leftarrow 1}(\hat{\rho}_1)=\prod_i\sum_m\left(\hat{K}_{m,i}\hat{\rho}_1\hat{K}^\dagger_{m.i}\right).
\end{equation}
Let us now construct the \textit{Kraus} operators $\hat{K}_i$. First, our choice of $\hat{K}_i$ must satisfy
\begin{equation}
    \sum_m\hat{K}^\dagger_{m,i}\hat{K}_{m,i}=\mathds{1}_i.
\end{equation}
This can be done in the following way. Note that we are able to decompose the coefficients of $\hat{\rho}_1$ and $\hat{X}_i\hat{\rho}_1\hat{X}_i$. That is,
\begin{equation}
\label{eq: param1}
\frac{1}{2}\left[1+e^{-\frac{2\lambda^2}{\omega^2}}\cos\theta\right] = e^{-\frac{2\lambda^2}{\omega^2}}\left(\frac{1+\cos\theta}{2}\right)+\frac{1-e^{-\frac{2\lambda^2}{\omega^2}}}{2} = a^2_1+a^2_2
\end{equation}
and
\begin{equation}
\label{eq: param2}
    \frac{1}{2}\left[1-e^{-\frac{2\lambda^2}{\omega^2}}\cos\theta\right] =e^{-\frac{2\lambda^2}{\omega^2}}\left(\frac{1-\cos\theta}{2}\right)+\frac{1-e^{-\frac{2\lambda^2}{\omega^2}}}{2}=b^2_1+b^2_3.
\end{equation}
Where we defined $\theta = 4\lambda p /\omega$. We then introduce three Kraus operators
\begin{equation}
\begin{aligned}
    \hat{K}_{1,i} = &a_1\hat{I}_i-ib_1\hat{X}_i\\
    \hat{K}_{2,i} = &a_2\hat{I}_i\\
    \hat{K}_{3,i} =& b_3\hat{X}_i
    \end{aligned}
\end{equation}
We confirm that these choice of Kraus operators indeed satisfy the required constraint
\begin{equation}
\sum^{3}_{m=1}\hat{K}^\dagger_{m,i}\hat{K}_{m,i} =\left( a^2_1+b^2_1+a^2_2+b^2_3\right)\hat{I}_i = \hat{I}_i.
\end{equation}
We therefore have
\begin{equation}
\begin{aligned}
\sum^3_{m=1}\hat{K}_{m,i}\hat{\rho}_1\hat{K}^\dagger_{m,i} =& \left(a_1\hat{I}_i-ib_1\hat{X}_i\right)\hat{\rho}_1\left(a_1\hat{I}_i+ib_1\hat{X}_i\right)+a^2_2\hat{\rho}_1+b^2_3\hat{X}_i\hat{\rho}_1\hat{X}_i\\
=&a^2_1\hat{\rho}_1+b^2_1\hat{X}_i\hat{\rho}_1\hat{X}_i-ia_1b_1\left[\hat{X}_i,\hat{\rho}_1\right]+a^2_2\hat{\rho}_1+b^2_3\hat{X}_i\hat{\rho}_1\hat{X}_i\\
=&\left(a^2_1+a^2_2\right)\hat{\rho}_1+\left(\hat{b}^2_1+\hat{b}^2_3\right)\hat{X}_i\hat{\rho}_1\hat{X}_i-ia_1b_1\left[\hat{X}_i,\hat{\rho}_1\right].
\end{aligned}
\end{equation}
Plugging in our parametrization given in Eq.~\eqref{eq: param1} and Eq.~\eqref{eq: param2},
\begin{equation}
\begin{aligned}
\sum^3_{m=1}\hat{K}_{m,i}\hat{\rho}_1\hat{K}^\dagger_{m,i} =&\frac{1}{2}\left[1+e^{-\frac{2\lambda^2}{\omega^2}}\cos\theta\right] \hat{\rho}_1+\frac{1}{2}\left[1-e^{-\frac{2\lambda^2}{\omega^2}}\cos\theta\right]\hat{X}_i\hat{\rho}_1\hat{X}_i-ie^{-\frac{\lambda^2}{\omega^2}}\sqrt{\frac{1+\cos\theta}{2}}e^{-\frac{\lambda^2}{\omega^2}}\sqrt{\frac{1-\cos\theta}{2}}\left[\hat{X}_i,\hat{\rho}_1\right] \\
=&\frac{1}{2}\left[1+e^{-\frac{2\lambda^2}{\omega^2}}\cos\theta\right] \hat{\rho}_1+\frac{1}{2}\left[1-e^{-\frac{2\lambda^2}{\omega^2}}\cos\theta\right]\hat{X}_i\hat{\rho}_1\hat{X}_i-\frac{ie^{-\frac{2\lambda^2}{\omega^2}}}{2}\sin\theta\left[\hat{X}_i,\hat{\rho}_1\right].
\end{aligned}
\end{equation}
Hence, we recover our original channel, Eq.~(\ref{eq:Channel}). We end this Appendix by categorizing the three Kraus operators as follows
\begin{equation}
\begin{cases}
    \hat{K}_{1,i} = e^{-\frac{\lambda^2}{\omega^2}}\sqrt{\frac{1+\cos(\frac{4\lambda p}{\omega})}{2}}\hat{I}_i-ie^{-\frac{\lambda^2}{\omega^2}}\sqrt{\frac{1-\cos(\frac{4\lambda p}{\omega})}{2}}\hat{X}_i\\
    \hat{K}_{2,i} =  \frac{1-e^{-\frac{2\lambda^2}{\omega^2}}}{2}\hat{I}_I\\
     \hat{K}_{3,i} = \frac{1-e^{-\frac{2\lambda^2}{\omega^2}}}{2}\hat{X}_i
    \end{cases}
\end{equation}
That is, upon choosing the bosonic coherent state as the variational wave function for bath phonons, the local error channel is composed of equal probabilities of 
doing nothing ($\hat K_2$), suffering a bit-flip error ($\hat K_3$), or collecting an error due to a rotation around the $\hat{X}_i$-axis ($\hat K_1$). 

\section{Tensor network methods for CMI calculations}
\label{sec: tensor network}

We extend the tensor network method developed in Sang et al.~\cite{Sang_2024_QMI} to calculate the entropy of a cluster state. As mentioned in the main text, the unique ground state of $\hat{H}_S$ can be written in the form $\hat{\rho}^\text{(PBC)}_\text{SPT} = \prod_{v\in A} \frac{1 + \hat{A}_v}{2} \prod_{e\in B} \frac{1 + \hat{B}_e}{2}$.
We are interested in calculating quantum information quantities associated with a subregion of the cluster state $M$, which could be non-simply-connected (i.e., with a hole in the middle). Recall from the main text that the reduced density matrix of the cluster state subject to the $\hat{X}$ noise writes:
\begin{equation}
    \hat{\rho}_{p,M} = \prod_{i\in M}\mathcal{N}^X_i\left[\prod_{v\in M}\frac{1+\hat{A}_v}{2}\prod_{e\in M}\frac{1+\hat{B}_e}{2}\right],
\end{equation}
see Eq.~(\ref{eq: noisey rdm}), where we now write $p$ explicitly indicating the error rate.

The dephasing channel that we apply here can be thought of as follows. With a probability $p$, it applies a $\hat{X}$ operator on each qubit. When the $\hat{X}$ operator acts on a vertex, it flips the two neighboring $B$ operators, creating two `edge' excitations; similarly, an $\hat{X}$ operator acting on an edge could create two `vertex' excitations. The key observation here is that edge excitations do not interfere with vertex excitations. Let us then use vectors ${\bf e}_{A}$, ${\bf e}_{B}$ to denote the noise configuration on the physical lattice that damages the information in $A$, $B$, and ${\bf m}_{A}$, ${\bf m}_{B}$ to denote the (vertex- and edge-) excitation configurations due to the noise. It is then convenient to write the corrupted reduced density matrix as a sum of excitation configurations
\begin{equation}
    \hat{\rho}_{p,M} \propto\prod_{{\bf m}_{A}}\Pr({\bf m}_{A})\Pi_{{\bf m}_{A}}\prod_{{\bf m}_{B}}\Pr({\bf m}_{B})\Pi_{{\bf m}_{B}},
\end{equation}
where $\Pr({\bf m}_{A, B}) = \sum_{{\bf{e}}_{A,B}}p^{{\bf e}_{A,B}}(1-p)^{{M_{A,B} -|{\bf e}_{A,B}|}}\delta(\partial{\bf e}_{A,B} = {\bf m}_{A,B})$. Here $\partial{\bf e}_{A, B} = {\bf m}_{A, B}$ corresponds to a noise configuration giving a certain excitation configuration. 
$\Pi_{{\bf m}_{A,B}}$ is the projector on a specific excitation configuration~\cite{Sang_2024_QMI}. 
The nth Renyi entropy can be calculated as
\begin{align}
    S^{(n)}(M) = \frac{1}{1-n}\log\Tr \hat{\rho}_{p,M}^{n} = S^{(n)}(\hat{\rho}_{p,M}) + H^{(n)}({\bf m}_{A},{\bf m}_{B}) = S^{(n)}(\hat{\rho}_{p,M}) + H^{(n)}({\bf m}_{A}) + H^{(n)}({\bf m}_{B}).
\end{align}
Here, $H^{(n)}({\bf m}_{A,B})$ denotes the $n$th R\'enyi entropy defined on $A$- and $B$- type excitations and $S^{(n)}(\hat{\rho}_p,M)$ corresponding to $n$th R\'enyi entropy of an excitation-free configuration.

This CMI can then be calculated with a tensor network method similar to~\cite{Sang_2024_QMI}, where a 2D tensor network can be constructed representing the probability over excitation configuration $\Pr({\bf m}_{A, B})$,
\begin{align}
    H^{(n)}({\bf m}_{A,B}) = \frac{1}{1-n}\log (\sum_{{\bf m}_{A,B}}[\Pr({\bf m}_{A,B})]^n)
\end{align}
The above equation can be evaluated numerically-exactly by contracting $n$ copies of the 2D tensor networks. For example, in the two-copy case, and for the calculation of A-type excitations, one could first append an additional `excitation leg' to each vertex tensor representing whether an excitation is created, then contract two copies of these 2D tensor networks.

At the scale we study in this work (up to $\sim92\times92$), exact contractions can be extremely hard to perform. Nevertheless, a `boundary contraction' method can be used~\cite{Murg_2007}
\begin{enumerate}
    \item Take the first row of a 2D tensor network as a matrix product state (MPS)
    \item Apply the second row of tensors as a matrix product operator (MPO) to the MPS. This will double the bond dimension of the original MPS.
    \item Perform a singular value decomposition (SVD) through the MPS and keep only the significant Schmidt values. 
    \item Repeat step 2 until all rows are contracted. 

\end{enumerate}
We numerically find that the contraction time scales roughly linearly with system size. This means that the structure of the problem guarantees that bond dimension does not increase as we perform row contractions.

\end{widetext}

\bibliography{bibliography}
\end{document}